\documentclass{article}

\usepackage{microtype}
\usepackage{graphicx}
\usepackage{caption}
\usepackage{subcaption}
\usepackage{booktabs} % for professional tables

\usepackage[table]{xcolor}

\usepackage{hyperref}

\usepackage[accepted]{icml2024}

\usepackage{amsmath}
\usepackage{amssymb}
\usepackage{mathtools}
\usepackage{amsthm}
\usepackage{tabularray}

\usepackage{svg}
\usepackage[utf8]{inputenc}
\usepackage{textcomp}
\usepackage{multirow}

\usepackage[capitalize,noabbrev]{cleveref}

\theoremstyle{plain}
\newtheorem{theorem}{Theorem}[section]

\newtheorem{lemma}[theorem]{Lemma}

\theoremstyle{definition}

\theoremstyle{remark}

\usepackage[textsize=tiny]{todonotes}

\newtoggle{comments}
\toggletrue{comments}
\iftoggle{comments}{
    \newcommand\AG[1]{\textcolor{magenta}{[AG: #1]}}
    \newcommand\TD[1]{\textcolor{cyan}{[TD: #1]}}
}{
    \newcommand\AG[1]{}
    \newcommand\TD[1]{}
}

\newcommand{\caduceus}{\includegraphics[height=0.7\baselineskip]{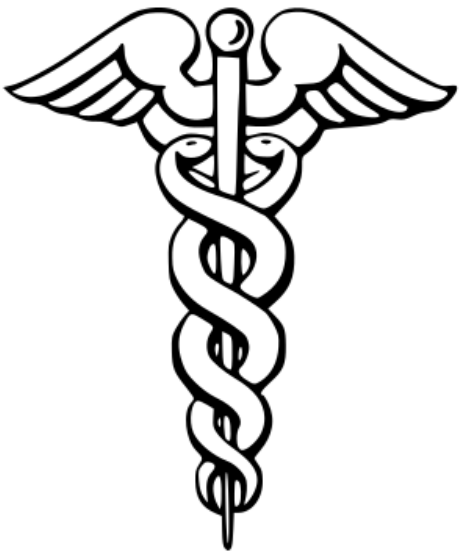}}
\definecolor{PhBlue}{RGB}{100,134,184}

\usepackage{amsmath,amsfonts,bm}

\def\Tabref#1{Table~\ref{#1}}
\def\Figref#1{Figure~\ref{#1}}

\def\Secref#1{Section~\ref{#1}}
\def\Appsecref#1{Appendix~\ref{#1}}
\def\eqref#1{equation~\ref{#1}}
\def\Eqref#1{Equation~\ref{#1}}
\def\Thrmref#1{Theorem~\ref{#1}}

\def\1{\bm{1}}

\def\rvx{{\mathbf{x}}}

\def\rmX{{\mathbf{X}}}

\def\vh{{\bm{h}}}

\def\vx{{\bm{x}}}
\def\vy{{\bm{y}}}

\def\mA{{\bm{A}}}
\def\mB{{\bm{B}}}
\def\mC{{\bm{C}}}
\def\mD{{\bm{D}}}

\def\mI{{\bm{I}}}

\def\mW{{\bm{W}}}

\DeclareMathAlphabet{\mathsfit}{\encodingdefault}{\sfdefault}{m}{sl}
\SetMathAlphabet{\mathsfit}{bold}{\encodingdefault}{\sfdefault}{bx}{n}

\def\sR{{\mathbb{R}}}

\newcommand{\RC}{\mathrm{RC}}
\newcommand{\M}{\mathrm{M}}
\newcommand{\concat}{\mathrm{concat}}
\newcommand{\splitchan}{\mathrm{split}}
\newcommand{\flipchan}{\mathrm{flip\_chan}}
\newcommand{\Emb}{\mathrm{Emb}}
\newcommand{\LM}{\mathrm{LM}}
\newcommand{\RCe}{\text{RCe}}

\icmltitlerunning{Caduceus: Bi-Directional Equivariant Long-Range DNA Sequence Modeling}

\begin{document}

\twocolumn[
\icmltitle{Caduceus: Bi-Directional Equivariant Long-Range DNA Sequence Modeling}

% \icmlsetsymbol{equal}{*}

\begin{icmlauthorlist}
\icmlauthor{Yair Schiff}{cornell}
\icmlauthor{Chia-Hsiang Kao}{cornell}
\icmlauthor{Aaron Gokaslan}{cornell}
\icmlauthor{Tri Dao}{princeton}
\icmlauthor{Albert Gu}{cmu}
\icmlauthor{Volodymyr Kuleshov}{cornell}
\end{icmlauthorlist}

\icmlaffiliation{cornell}{Department of Computer Science, Cornell University, New York, NY USA}
\icmlaffiliation{cmu}{School of Computer Science Carnegie Mellon University, Pittsburgh, PA USA}
\icmlaffiliation{princeton}{Department of Computer Science, Princeton University, Princeton, NJ USA}

\icmlcorrespondingauthor{Yair Schiff}{\href{mailto://yzs2@cornell.edu}{\texttt{yzs2@cornell.edu}}.}

\icmlkeywords{Machine Learning, Deep Learning, Language Modeling, Foundation Models, Genomics}

\vskip 0.3in
]

% this must go after the closing bracket ] following \twocolumn[ ...

% This command actually creates the footnote in the first column
% listing the affiliations and the copyright notice.
% The command takes one argument, which is text to display at the start of the footnote.
% The \icmlEqualContribution command is standard text for equal contribution.
% Remove it (just {}) if you do not need this facility.

%\printAffiliationsAndNotice{}  % leave blank if no need to mention equal contribution
\printAffiliationsAndNotice{}
% \icmlEqualContribution} % otherwise use the standard text.

\begin{abstract}
Large-scale sequence modeling has sparked rapid advances that now extend into biology and genomics. However, modeling genomic sequences introduces challenges such as the need to model long-range token interactions, the effects of upstream and downstream regions of the genome, and the reverse complementarity (RC) of DNA. Here, we propose an architecture motivated by these challenges that builds off the long-range Mamba block, and extends it to a BiMamba component that supports bi-directionality, and to a MambaDNA block that additionally supports RC equivariance. We use MambaDNA as the basis of Caduceus, the first family of RC equivariant bi-directional long-range DNA language models, and we introduce pre-training and fine-tuning strategies that yield Caduceus DNA foundation models. Caduceus outperforms previous long-range models on downstream benchmarks; on a challenging long-range variant effect prediction task, Caduceus exceeds the performance of 10x larger models that do not leverage bi-directionality or equivariance. Code to reproduce our experiments is available \href{https://github.com/kuleshov-group/caduceus}{here}.
\end{abstract}
\begin{figure*}[ht!]
    \centering
    \includegraphics[width=1.8\columnwidth]{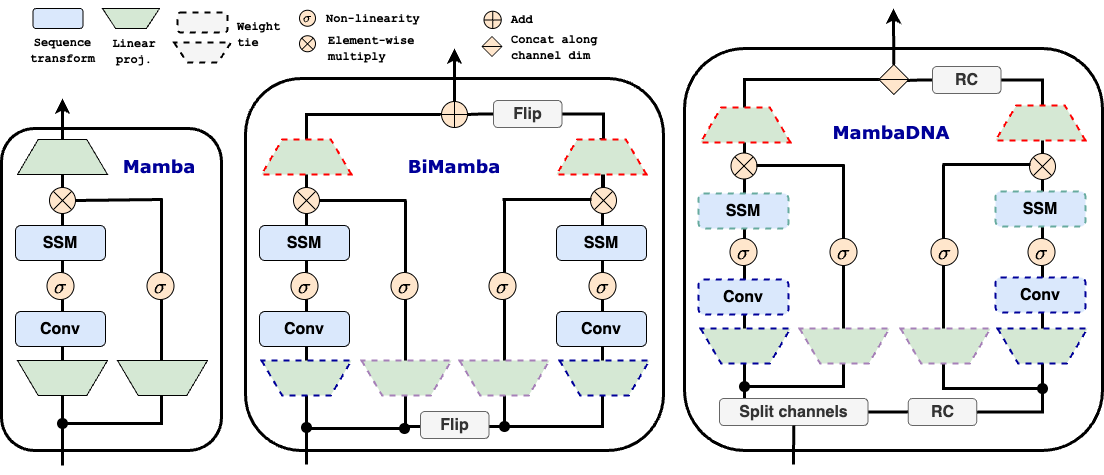}
    \caption{
        Mamba modules for genomic sequences.
        \textit{(Left)} \textbf{Mamba}: The original left-to-right causal Mamba module proposed in \citet{gu2023mamba}.
        \textit{(Middle)} \textbf{BiMamba}: A parameter efficient bi-directional extension of the Mamba module.
        In-projection and out-projection parameters are shared for processing the sequence and its reverse.
        After processing the reversed sequence, it is flipped again and added to the forward output.
        \textit{(Right)} \textbf{Reverse complementary equivariant Mamba (MambaDNA)}: A module with RC equivariance inductive bias.
        The input is first split into two along the channel dimension.
        One split has the reverse complement (RC) operation applied to it.
        All the parameters of a Mamba module are shared for processing the forward and RC sequence.
        The reverse sequence has the RC applied once more before being concatenated back with the forward output along the channel dimension.
    }
    \label{fig:mamba}
\end{figure*}

\section{Introduction}\label{sec:intro}
Large-scale sequence models have sparked rapid progress in machine learning, bringing about advances that extend beyond natural language processing (NLP) \citep{achiam2023gpt, team2023gemini} into science, biology, and medicine.
In proteomics, these models have enabled predicting protein structures from sequences \citep{jumper2021highly,lin2023evolutionary}, deciphering the functions and interactions of amino acids \citep{rao2020transformer, rives2021biological}, and crafting new molecules \citep{madani2023large}. As compute cost decreases, sequence modeling is poised to further impact biology.

Sequence models are also standard tools in genomics
% , e.g., for tasks such as predicting the outcome of functional assays and the effects of genetic mutations on cellular phenotypes 
\citep{zhou2015predicting, avsec2021effective}. 
Unlike proteins, genomes contain non-coding sequences, which often play an important role in regulating cellular mechanisms, and can thus potentially provide greater insights into cell biology.
Understanding non-coding sequences has been a key focus of recent work, including efforts in applying large language models (LMs) to genomes \citep{ji2021dnabert, benegas2023gpn, dalla2023nucleotide, nguyen2023hyenadna}.

However, modeling DNA introduces challenges that are distinct from those posed by natural language or proteins.
First, cellular phenotypes are often impacted by base pairs both upstream and downstream in the genome, which requires sequence models to handle bi-directional context.
Second, DNA consists of two strands that are reverse complements of each other and that carry the same information; modeling this property can significantly improve performance \citep{zhou2021towards, mallet2021reverse}.
Third, many genomics tasks, such as predicting the effect of variants on gene expression, can entail long-range interactions, as nucleic acids even up to 1 million base pairs away from a gene can have significant regulatory effects \citep{furlong2018_enhancers}.

In this paper, we propose architectural components motivated by the above challenges.
Our modules build off the long-range Mamba block \citep{gu2023mamba} and thus naturally handle long sequences of over hundreds of thousands of nucleotides without the quadratic cost of attention-based architectures \citep{vaswani2017attention}.
We extend Mamba to BiMamba, a component that supports bi-directionality, and to MambaDNA, which further adds reverse complement (RC) equivariance.
The MambaDNA block can be used as a drop-in replacement in architectures for genome analysis in both supervised and self-supervised contexts.

We then use MambaDNA as the basis of Caduceus\footnote{Caduceus (\caduceus) is the staff carried by Hermes in Greek mythology that is adorned by two intertwined serpents.
We choose this name to evoke imagery of the double helix structure of DNA
% , an association with medicine,
and to symbolize bi-directionality using a Mamba sequence operator.}, a family of bidirectional long-range DNA sequence models that is the first to support RC equivariant language modeling.
We further introduce pre-training and fine-tuning strategies that yield Caduceus foundation models for a wide range of predictive tasks in genomics.
The Caduceus models consistently outperform previous SSM-based models of a similar size.
% \citep{nguyen2023hyenadna}
% in terms of 
% pre-training loss and 
% downstream performance. % in large part due to their ability to incorporate bi-directionality and RC-equivariance. 
On many tasks, especially ones that require long-range modeling, Caduceus also outperforms 10x larger Transformer-based models.
% \citep{dalla2023nucleotide}.

We use Caduceus to perform variant effect prediction (VEP), a task that seeks to determine whether a genetic mutation influences a phenotype---gene expression in our case.
This task is a natural fit for Caduceus because its pre-training implicitly learns to recognize the effects of evolutionary pressure (e.g., conservation, co-evolution), which is a key source of signal for VEP (e.g., a mutation in a region where mutations are rare likely has an effect and a low probability under the model).
On a task derived from a standard dataset of mutations with long-range effects on gene expression \citep{avsec2021effective}, Caduceus outperforms existing attention and SSM-based models that do not leverage both bi-directionality and equivariance.
\paragraph{Contributions}
To summarize, our contributions are:
\begin{enumerate}
    \item We introduce BiMamba, a parameter and hardware efficient extension of the Mamba block that supports bi-directional sequence modeling.
    \item We extend BiMamba to support RC equivariance, which yields the MambaDNA block, a general component for deep learning architectures in genomics.
    \item We use MambaDNA as the basis of Caduceus, the first family of RC-equivariant DNA foundation models.
    \item We demonstrate that on long-range tasks, Caduceus outperforms models that are up to 10x larger but that do not use bi-directionality or equivariance.
\end{enumerate}

\section{Background}\label{sec:background}
\subsection{DNA Terminology}\label{subsec:dna}
\textit{Deoxyribonucleic acid} (DNA) is a polymer that is made up of two complementary strands that wind in a ladder / double-helix manner and is comprised of four \textit{nucleotide} bases:  \textit{adenine} (\texttt{A}), \textit{cytosine} (\texttt{C}), \textit{guanine} (\texttt{G}) or \textit{thymine} (\texttt{T}).
The bonds between the nucleotide bases form `rungs' on the twisted ladder, with \texttt{A} bonding with \texttt{T} and \texttt{C} bonding with \texttt{G}.
DNA contains the genetic code for forming proteins.
In complex organisms, DNA can be billions of nucleotide base pairs (bps) long, but the long strands coil tightly around proteins in the nucleus called \textit{histones}.

Genetic mutations at individual bps, known as \textit{single nucleotide polymorphisms} (SNPs) can account for phenotypic variation across organisms.
Evolutionary pressure has forced several genomic regions to be conserved across time and species, with deleterious mutations failing to proliferate in populations.
Mutations in conserved regions can therefore have an out-sized effect on phenotype, and models that can identify these regions will likely perform better on variant effect prediction tasks.

\paragraph{Reverse Complement Strands}
In the double-helix DNA structure, each strand contains semantically equivalent information.
The `reverse complement' (RC) of a given strand is oriented in the opposite direction of its counterpart with bps complemented relative to the `forward' strand: \texttt{A} converted to \texttt{T} and \texttt{C} to \texttt{G}.
In many biological assays, either strand of the DNA can be sequenced with equal probability.
However, learning to recognize non-palindromic DNA sequence motifs can be difficult for standard models \citep{zhou2021towards}.
Therefore, enforcing RC equivariance, loosely defined as model outputs transforming in a manner commensurate with RC-ing an input sequence, is an important desiderata of DNA sequence modeling.

\subsection{Structured State Space Models}\label{subsec:ssm}
A recent class of sequence models known as Structured State Space Models (SSMs\footnote{The acronym SSM is commonly used in machine learning communities to refer to this class of models, while in other disciplines it is typically associated to the broader class of state space models widely used in engineering.}; \citet{gu2021efficiently, gu2021combining,gu2022parameterization, gupta2022diagonal, smith2022simplified, dao2022hungry}) have proven to be effective at handling long-range models.
At the core of all of these models is a pair of linear differential equations that govern the mapping from input sequences $x(t) \in \sR$ to output sequences $y(t) \in \sR$ through an intermediate representation $\vh(t) \in \sR^N$:
\begin{align}\label{eq:ssm}
\begin{split}
    \dot{\vh}(t) = \mA h(t) + \mB x(t), ~~~~ y(t) = \mC h(t) + \mD x(t),
\end{split}
\end{align}
where $\mA \in \mathbb{R}^{N \times N}, \mB \in \mathbb{R}^{N \times 1}, \mC \in \mathbb{R}^{1 \times N},$ and $\mD \in \mathbb{R}$ are the parameters of the system.
For multidimensional sequences, $\vx(t), \vy(t) \in \sR^D,$ these dynamics are applied independently to each component.

This differential equation can be discretized with the continuous parameters converted, as follows:
\begin{align}\label{eq:ssm_discrete}
\begin{split}
    \vh_{t+1} = \overline{\mA}\vh_t + \overline{\mB}x_t, ~~~~ y_{t+1} = \mC \vh_t + \mD x_t,
\end{split}
\end{align}
by means of some discretization formula that is a function of continuous parameters $\mA, \mB,$ and an additional time scale parameter $\Delta.$
% Several popular discretization schemes exist, but
A common discretization used in the SSM literature is the zero-order hold, defined as:
\begin{align}\label{eq:zoh}
    \overline{\mA} = \exp(\Delta \mA), ~~~~
    \overline{\mB} = \mA^{-1}(\exp(\Delta \mA) - \mI)\mB.
\end{align}
Importantly, the linear-time invariance (LTI) of \Eqref{eq:ssm} allows us to equivalently formulate \Eqref{eq:ssm_discrete} as a convolution by unrolling the recurrence, enabling efficient parallel computation during training.

\paragraph{Selection Mechanisms}
However, the computational efficiency of the LTI formulation comes at the cost of the model not being able to adapt / attend to specific inputs.
To alleviate this lack of expressivity, \citet{gu2023mamba} introduce a \textit{selective} SSM that enables dependence of the parameters $\mB, \mC,$ and $\Delta$ on the input $x(t),$ with:
\begin{align}\label{eq:selection}
\begin{split}
    \mB_t = &\mathrm{Linear}_{\mB}(x_t) ~~~~~~
    \mC_t = \mathrm{Linear}_{\mC}(x_t) \\ %\nonumber
    &\Delta_t = \mathrm{softplus}(\mathrm{Linear}_{\Delta}(x_t)),
\end{split}
\end{align}
where $\mathrm{Linear}(\cdot)$ represents a linear projection and $\mathrm{softplus}(\cdot) = \log(1 + \exp(\cdot)).$

While this formulation renders $\overline{\mA}_t$ and $\overline{\mB}_t$ time-dependent, the linear recurrence in \Eqref{eq:ssm_discrete} can be formulated as an associative scan \citep{martin2017parallelizing}, which allows us to use an efficient parallel algorithm \citep{blelloch1990} and reduce computation to a logarithmic in sequence length.

\paragraph{Mamba}
The Mamba block presented in \citet{gu2023mamba} is formed by combining a selective SSM sequence transformation and a gated MLP mechanism.
This is depicted in the left-most schematic in \Figref{fig:mamba}.
An incoming sequence is copied and projected to twice the input dimension.
One copy is then passed through a causal convolution, followed by the SiLU/Swish non-linear activation \citep{ramachandran2017searching} and then finally through the selective SSM.
The other copy has the SiLU non-linearity applied to it and then gates the SSM output.
The gated representation is then projected back to the original dimension $D.$
As this is a causal, left-to-right sequence operation, the original models that use Mamba blocks are trained with the next token prediction (NTP) objective during pre-training.

\section{Bi-Directional \& RC-Equivariant Mamba}\label{sec:mamba_ext}
In this section, we present components that extend the Mamba block \citep{gu2023mamba}.
While these extensions are domain-agnostic, they are relevant to modeling DNA.

\subsection{BiMamba}\label{subsec:mamba_bidir}
The first extension that we apply to the standard Mamba module is to convert it from causal (left-to-right) to bi-directional. 
We achieve this by applying the Mamba module twice: once to the original sequence and once to a copy that is reversed along the length dimension.
To combine information, the output of the reversed sequence is flipped along the length dimension and added to the forward one.

A naive implementation of this method would double the number of parameters of the module.
To avoid this added memory footprint, we instead share projection weights between the `forward' and `reverse' Mamba.
These projections account for a vast majority of the model's parameters compared to those in the convolution and the SSM sub-modules \cite{gu2023mamba}.
We refer to this parameter efficient bi-directional block as \textbf{BiMamba}.
This module is depicted in the middle schematic of \Figref{fig:mamba}.

\subsection{MambaDNA}\label{subsec:mamba_rc}
To encode the RC equivariance inductive bias into our modules, we apply a Mamba (or BiMamba) block to a sequence and its RC, with parameters shared between the two applications \citep{shrikumar2017reverse, zhou2021towards}.
Given its relevance to genomics, we dub this block \textbf{MambaDNA}.

Concretely, let $\rmX_{1:T}^{1:D}$ denote a sequence of length $T$ with $D$ channels.
The channel splitting operation is then defined as:
\begin{align*}
    \splitchan(\rmX_{1:T}^{1:D}) := \left[\rmX_{1:T}^{1:(D/2)}, \rmX_{1:T}^{(D/2):D}\right].
\end{align*}
We also define the RC operation as follows:
\begin{align*}
    \RC\left(\rmX_{1:T}^{1:D}\right) := \rmX_{T:1}^{D:1}
\end{align*}
Finally, letting $\concat$ denote the last operation of this module that re-combines the sequences along the channel dimension, our RC equivariant Mamba module, which we denote as $\M_{\RCe, \theta},$ can be expressed as follows:
\begin{align*}
\begin{split}
    &\M_{\RCe, \theta}\left(\rmX_{1:T}^{1:D}\right) :=\\
    &\concat\left(\left[
    \M_\theta\left(\rmX_{1:T}^{1:(D/2)}\right), \RC\left(\M_\theta\left(\rmX_{T:1}^{D:(D/2)}\right)\right)\right]\right),
\end{split}
\end{align*}
where $\M_\theta$ represents the sequence operator that is parameterized by either the standard Mamba or BiMamba.
The MambaDNA module is depicted in the rightmost schematic of \Figref{fig:mamba}, with $\M_\theta$ shown as the standard Mamba.

We claim that MambaDNA satisfies the RC equivariance property that we desire for processing DNA sequences:
\begin{theorem}\label{thrm:mamba_rcequiv}
    The $\M_{\RCe,\theta}$ operator satisfies the following:
    \begin{align*}
    \RC \circ \M_{\RCe,\theta}\left(\rmX_{1:T}^{1:D}\right) = 
    \M_{\RCe,\theta} \circ \RC\left(\rmX_{1:T}^{1:D}\right).
    \end{align*}
\end{theorem}
\textit{Proof.} See \Appsecref{appsec:mamba_rcequiv_proof}.

Similar to BiMamba modules, MambaDNA blocks do not entail significant additional memory footprint, since the wrapped sequence operator that processes the forward and RC sequences is completely shared.

\begin{figure}[ht!]
    \centering
    \includegraphics[width=0.9\columnwidth]{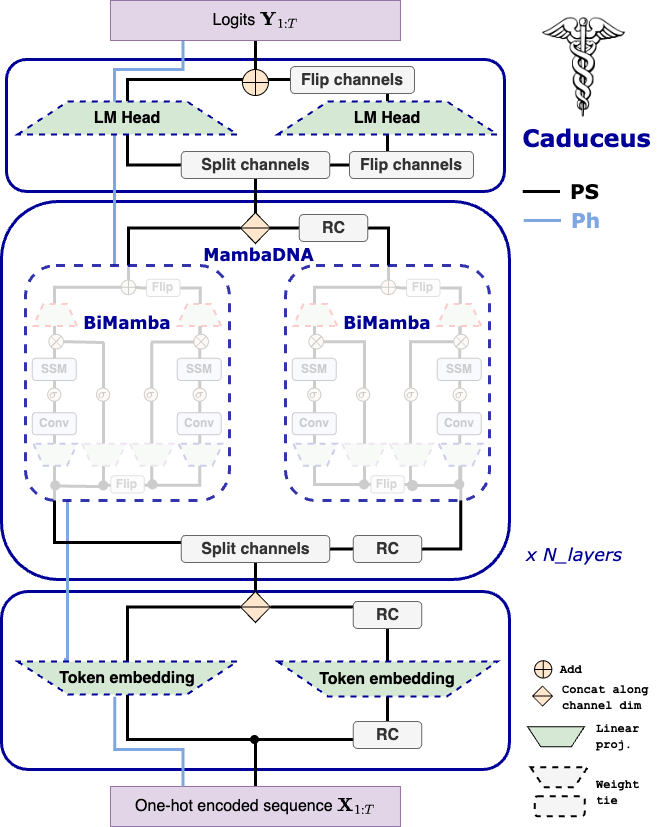}
    \caption{Caduceus Architecture.
    Bi-directional, RC equivariant Mamba modules are used in conjunction with equivariant word embeddings and language model head to form \textbf{Caduceus-PS}.
    Using only BiMamba blocks with RC data augmentation during pretraining and post-hoc conjoining for downstream task inference yields \textbf{\textcolor{PhBlue}{Caduceus-Ph}}.
    \scriptsize{\textsc{Caduceus Image license: Creative Commons CC0 1.0 Universal Public Domain Dedication.}
    }}
    \label{fig:caduceus}
\end{figure}
\section{Caduceus~\caduceus}\label{sec:caduceus}
Below we describe \textbf{Caduceus}, a novel bi-directional DNA LM architecture that enforces RC equivariance.
We introduce two versions of this model, each of which maintains equivariance in a different manner: either (1) via parameter sharing \citep{shrikumar2017reverse}, Caduceus-PS, or (2) via a technique used during downstream task inference, known as \textit{post-hoc conjoining} \citep{zhou2021towards}, Caduceus-Ph.

\subsection{Caduceus-PS}\label{subsec:caduceus_ps}
\paragraph{Architecture}
For \textbf{Caduceus-PS}, we leverage both of the architectural innovations introduced in \Secref{sec:mamba_ext}.
Namely, we wrap a BiMamba module within a MambaDNA block.
Additionally, preceding the Mamba blocks of this architecture is an RC equivariant token embedding module.
Denoting by $\Emb_\theta$ the linear projection that takes one-hot vectors $\rmX_{1:T}^{1:4}$ and produces embeddings in $\sR^{D/2},$ the RC equivariant version of this embedding is defined as:
\begin{align*}
\begin{split}
    &\Emb_{\RCe,\theta}\left(\rmX_{1:T}^{1:4}\right) := \\
    &\concat\left(\left[\Emb_\theta\left(\rmX_{1:T}^{1:4}\right), \RC\circ\Emb_\theta\left(\RC\left(\rmX_{1:T}^{1:4}\right)\right)\right]\right)
\end{split}
\end{align*}
Additionally, the logits of the Caduceus model are produced by passing the output of its final MambaDNA block through a RC equivariant language model head.
To our knowledge, Caduceus-PS is the first model to incorporate RC equivariance into the LM pre-training paradigm.
This can be formalized by first defining a channel flip operator $\flipchan\left(\rmX_{1:T}^{1:D}\right) := \left(\rmX_{1:T}^{D:1}\right).$
Then, letting $\LM_\theta$ be the linear projection from sequences with $D/2$ channels to vectors in $\sR^4$, we define the equivariant version of the language modeling head as:
\begin{align*}
    &\LM_{\RCe, \theta}\left(\rmX_{1:T}^{1:D}\right) := \\
    &\LM_\theta\left(\rmX_{1:T}^{1:(D/2)}\right) + \flipchan\circ\LM_\theta\left(\rmX_{1:T}^{D:(D/2)}\right).
\end{align*}
Depicted in \Figref{fig:caduceus} with the \textbf{black} path, Caduceus-PS enables RC equivariant pre-training: the predictions it produces for the RC of a given sequence are equivalent to reversing the predictions of the original sequence along the length dimension and complementing outputs: \texttt{A}-\texttt{T} and \texttt{C}-\texttt{G}.
We formalize this claim in the following statement:
\begin{theorem}\label{thrm:caduceus_rcequiv}
    Composing $\LM_{\RCe, \theta} \circ \M_{\RCe, \theta}^{(n)} \circ \Emb_{\RCe, \theta}$, where $\M_{\RCe, \theta}^{(n)}$ denotes $n$ compositions of Mamba RC equivariant modules, yields an operator that is RC equivariant.
\end{theorem}
\textit{Proof.} See \Appsecref{appsec:caduceus_rcequiv_proof}.

\paragraph{Pre-training}
Given the bi-directionality of this model, we train Caduceus-PS with the masked language modeling (MLM) objective, using the standard masking recipe proposed in BERT \citep{devlin2018bert}.
The RC equivariant language modeling of Caduceus-PS means that we do not need RC data augmentation at pre-training, since predictions are inherently symmetric with respect to this operation.

\paragraph{Downstream Usage}
For downstream tasks, since either strand of an assayed sequence will carry the same label, we wish to enforce RC \textit{invariance}.
The token embedding parameter sharing in Caduceus-PS means that its intermediate and final hidden states are twice the (channel) dimensionality of a standard Mamba-based language model with an equivalently sized token embedding matrix.
To enforce RC invariance at downstream training and inference, final hidden states are split and the two splits are averaged.

\subsection{Caduceus-Ph}\label{subsec:caduceus_ph}

\paragraph{Architecture}
The \textbf{\textcolor{PhBlue}{Caduceus-Ph}} model is depicted with the \textbf{\textcolor{PhBlue}{blue}} path in \Figref{fig:caduceus}.
The core of this model is a stack of BiMamba blocks.

\paragraph{Pre-training}
As with Caduceus-PS, this model is pre-trained using the same MLM objective.
However, as the model is not an RC equivariant LM, we instead rely on data augmentation during pre-training.

\paragraph{Downstream Usage}
In order to make the downstream task representations RC invariant, we leverage a technique called \textit{post-hoc} conjoining \citep{zhou2021towards}.
Namely, for downstream task \textit{training} the backbone model is unchanged, but we employ RC data augmentation.
However, for downstream task \textit{inference}, we apply the model twice, once on the original sequence and once on a corresponding RC sequence, and average the two, effectively performing a version of `RC ensembling' \citep{mallet2021reverse}.

\section{Experiments}\label{sec:exp}
\subsection{Pre-training}\label{subsec:exp_pretrain}
\begin{figure*}[ht!]
    \centering
    \begin{subfigure}[c]{0.32\textwidth}
    \centering
    \includegraphics[width=\textwidth]{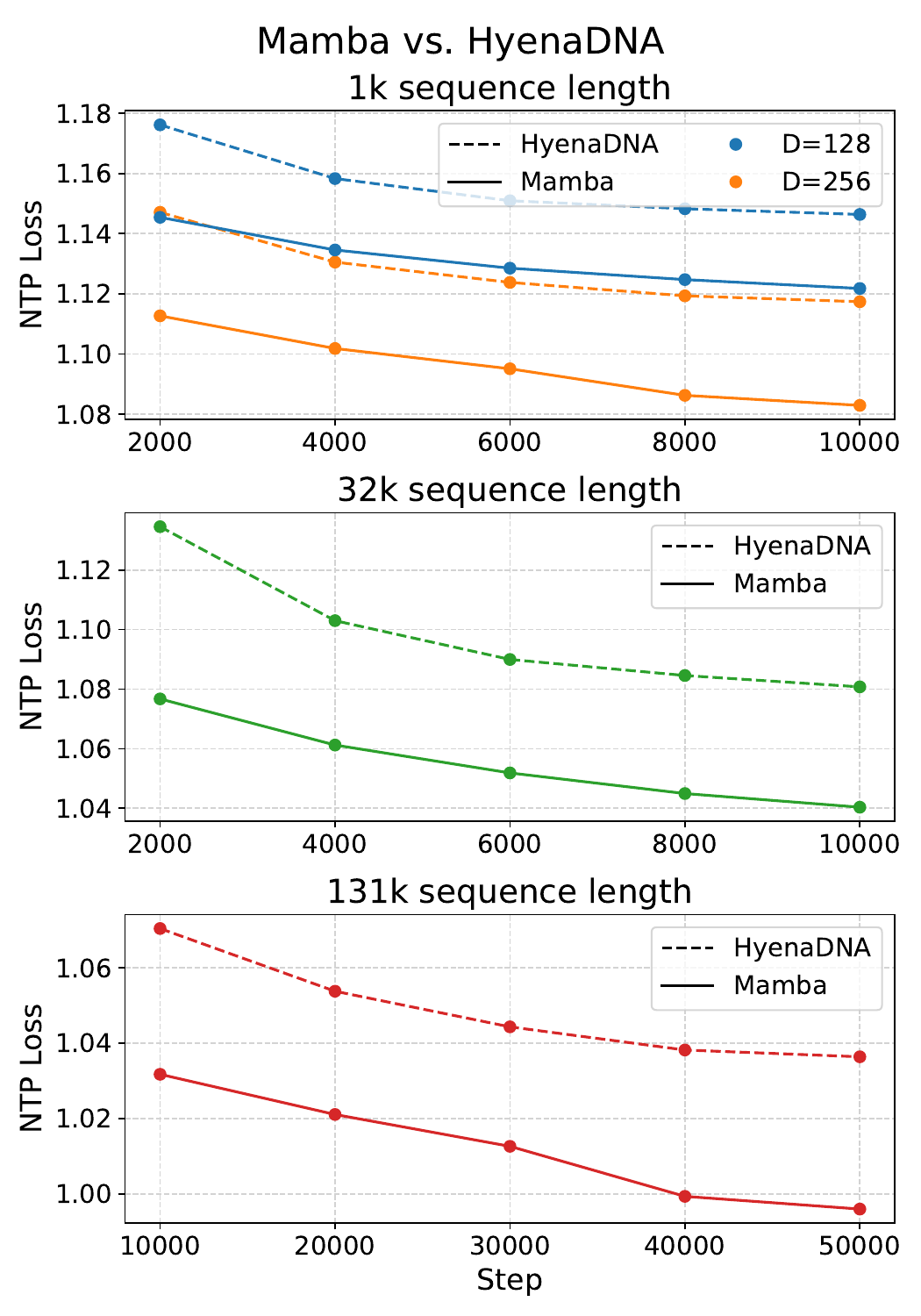}
    \caption{}
    \label{subfig:mamba_ntp_vs_hyenadna}
    \end{subfigure}
    \hfill
    \begin{subfigure}[c]{0.32\textwidth}
    \centering
    \includegraphics[width=\textwidth]{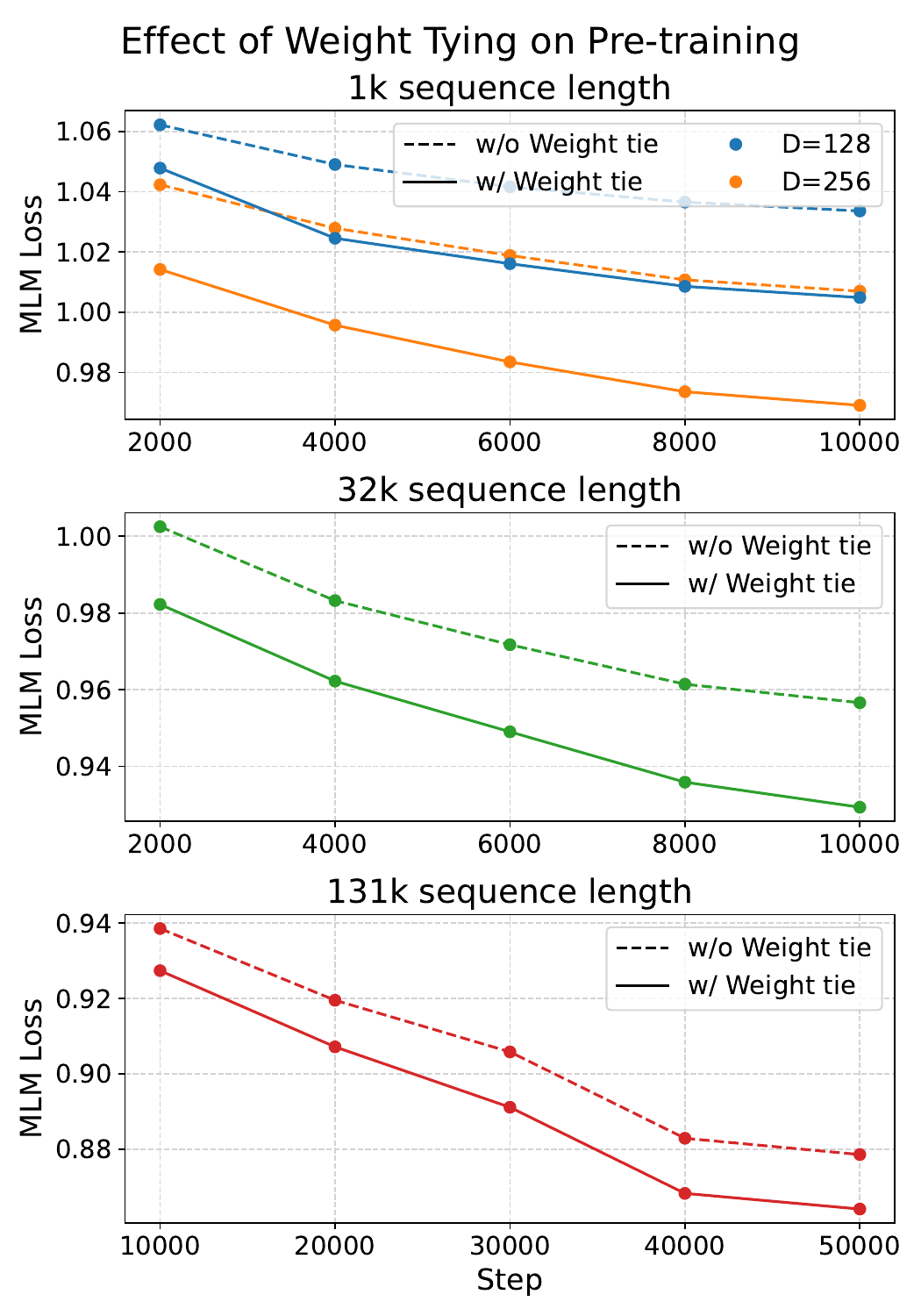}
    \caption{}
    \label{subfig:mamba_weight_tie}
    \end{subfigure}
    \hfill
    \begin{subfigure}[c]{0.32\textwidth}
    \centering
    \includegraphics[width=\textwidth]{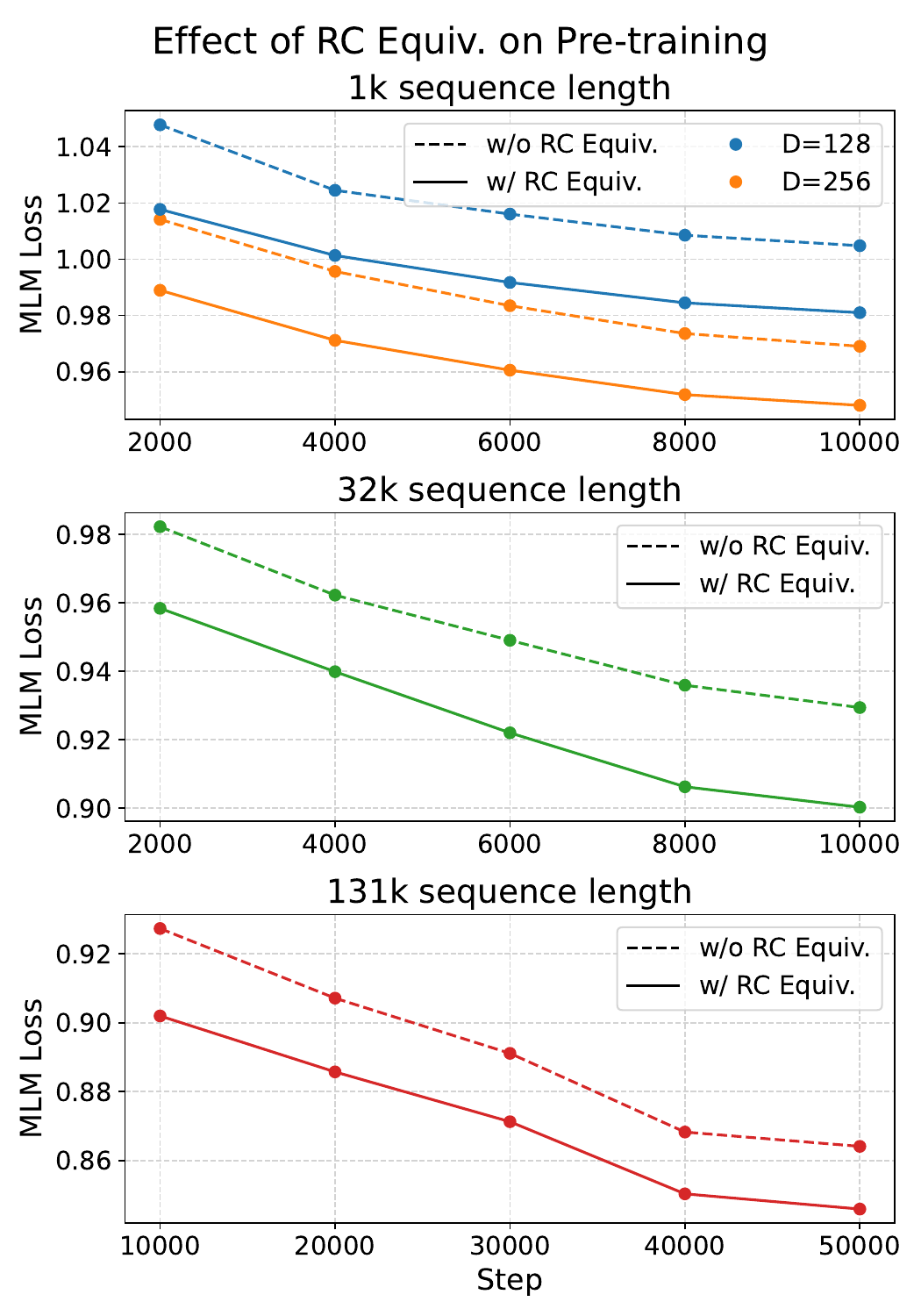}
    \caption{}
    \label{subfig:mamba_rc}
    \end{subfigure}
    \caption{
        Pre-training test set loss. 
        (a) For comparable model size and sequence length, Mamba attains better cross entropy loss than HyenaDNA during pre-training on the human genome.
        (b) Across sequence lengths, deeper models that use weight tying have better pre-training loss on the human genome.
        (c) Across sequence lengths, RC equivariance leads to better pre-training loss on the human genome.
        Note, models with a sequence length of 131k were validated less frequently to reduce overhead during pre-training.
        By adjusting batch size, we hold number of tokens per batch constant across varying lengths.
    }
    \label{fig:pre_train}
\end{figure*}
\paragraph{Data}
We limit the focus of this work to human-genome related tasks.
To that end, we perform all pre-training tasks on the human reference genome \citep{genome2009genome}.
% , which consists of approximately 4.5 billion bps.
We use character- / base pair-level tokenization.
While other DNA FMs have explored k-mer tokenization, this scheme suffers from the drawback that minor changes to an input sequence can lead to drastically different tokenization outputs~\citep{zhou2023dnabert}, which complicates training.
Character-level tokenization avoids this issue.
For any non-RC equivariant model that we train, including re-training HyenaDNA \citep{nguyen2023hyenadna} models, we employ RC data augmentation during pre-training.
For more information on the pre-training dataset and recipes see \Appsecref{appsec:pretrain}.

\paragraph{Mamba vs. HyenaDNA NTP}
Similar to the preliminary results in \citet{gu2023mamba}, we find that the Mamba module performs better than Hyena in terms of NTP.
In Figure \ref{subfig:mamba_ntp_vs_hyenadna}, we see that at varying sequence lengths and comparable model sizes, a standard Mamba model attains lower cross entropy loss compared to HyenaDNA.
As reported in \citet{gu2023mamba}, we also found that Mamba is more robust to higher learning rates, a common best practice in training LMs.
These results lend support to our choice of Mamba as the inner building block of our models.

\paragraph{Effect of Parameter Sharing on MLM Pre-training}
Projection parameter sharing in BiMamba enables deeper bi-directional models for similar parameter counts.
We compare MLM pre-training loss of BiMamba models to naive bi-directional Mamba models that do not use weight tying and are therefore reduced to half the depth.
We find that our parameter efficient implementation of bi-directionality leads to better pre-training loss, as seen in Figure \ref{subfig:mamba_weight_tie}.

\paragraph{Effect of RC Equivariance on MLM Pre-training}
We also examine the effect of using our proposed RC equivariant LM on pre-training.
In Figure \ref{subfig:mamba_rc}, we find that RC equivariant LM leads to better MLM pre-training loss.
This is significant because, as described above, performance on the MLM task has grounding in the biology of downstream tasks, such as variant effect prediction.

\subsection{Downstream Tasks}\label{subsec:downstream}

\subsubsection{Genomics Benchmark}\label{subsubsec:genomics_benchmark}

\begin{table*}[ht!]
\centering
\caption{Genomic Benchmarks. Top-1 accuracy ($\uparrow$) across 5-fold cross-validation (CV) for pretrained HyenaDNA, Mamba NTP, Caduceus models, and a supervised CNN baseline (trained from scratch).
Best values per task are \textbf{bolded}, second best are \textit{italicized}.
Error bars indicate the difference between the maximum and minimum values across 5 random seeds used for CV.
}
\label{tab:genomic_benchmarks}
% \vspace{-0.1in}
\begin{center}
\begin{scriptsize}
\begin{sc}
\begin{tabular}{lcccccc}
\toprule
& \begin{tabular}{@{}c@{}}CNN \\ (264k) \end{tabular} 
& \begin{tabular}{@{}c@{}}HyenaDNA \\ (436k) \end{tabular} 
& \begin{tabular}{@{}c@{}}Mamba\\ (468k) \end{tabular} 
& \begin{tabular}{@{}c@{}}Caduceus \\ w/o Equiv. \\ (470k) \end{tabular} 
& \begin{tabular}{@{}c@{}}Caduceus-Ph \\ (470k)\end{tabular}
& \begin{tabular}{@{}c@{}}Caduceus-PS \\ (470k) \end{tabular} \\
\midrule
Mouse Enhancers & $0.715$ \tiny{$\pm 0.087$} & $\it0.780$ \tiny{$\pm 0.025$} & $0.743$ \tiny{$\pm 0.054$} & $0.770$ \tiny{$\pm 0.058$} & $0.754$ \tiny{$\pm 0.074$} & $\bf0.793$ \tiny{$\pm 0.058$} \\
Coding vs. Intergenomic & $0.892$ \tiny{$\pm 0.008$} & $0.904$ \tiny{$\pm 0.005$} & $0.904$ \tiny{$\pm 0.004$} & $0.908$ \tiny{$\pm 0.003$} & $\bf0.915$ \tiny{$\pm 0.003$} & $\it0.910$ \tiny{$\pm 0.003$} \\
Human vs. Worm & $0.942$ \tiny{$\pm 0.002$} & $0.964$ \tiny{$\pm 0.002$} & $0.967$ \tiny{$\pm 0.002$} & $\it0.970$ \tiny{$\pm 0.003$} & $\bf0.973$ \tiny{$\pm 0.001$} & $0.968$ \tiny{$\pm 0.002$} \\
Human Enhancers Cohn & $0.702$ \tiny{$\pm 0.021$} & $0.729$ \tiny{$\pm 0.014$} & $0.732$ \tiny{$\pm 0.029$} & $0.741$ \tiny{$\pm 0.008$} & $\bf0.747$ \tiny{$\pm 0.004$} & $\it0.745$ \tiny{$\pm 0.007$} \\
Human Enhancer Ensembl & $0.744$ \tiny{$\pm 0.122$} & $0.849$ \tiny{$\pm 0.006$} & $0.862$ \tiny{$\pm 0.008$} & $0.883$ \tiny{$\pm 0.002$} & $\it0.893$ \tiny{$\pm 0.008$} & $\bf0.900$ \tiny{$\pm 0.006$} \\
Human Regulatory & $0.872$ \tiny{$\pm 0.005$} & $0.869$ \tiny{$\pm 0.012$} & $0.814$ \tiny{$\pm 0.211$} & $0.871$ \tiny{$\pm 0.007$} & $\it0.872$ \tiny{$\pm 0.011$} & $\bf0.873$ \tiny{$\pm 0.007$} \\
Human OCR Ensembl & $0.698$ \tiny{$\pm 0.013$} & $0.783$ \tiny{$\pm 0.007$} & $0.815$ \tiny{$\pm 0.002$} & $0.818$ \tiny{$\pm 0.003$} & $\bf0.828$ \tiny{$\pm 0.006$} & $\it0.818$ \tiny{$\pm 0.006$} \\
Human NonTATA Promoters & $0.861$ \tiny{$\pm 0.009$} & $0.944$ \tiny{$\pm 0.002$} & $0.933$ \tiny{$\pm 0.007$} & $0.933$ \tiny{$\pm 0.006$} & $\bf0.946$ \tiny{$\pm 0.007$} & $\it0.945$ \tiny{$\pm 0.010$} \\

\bottomrule
\end{tabular}
\end{sc}
\end{scriptsize}
\end{center}
\vskip -0.1in
\end{table*}

% \begin{table*}[th]
% \centering
% \begin{tabular}{lcccccc}
% \toprule
% {Dataset} & {CNN} & {HyenaDNA} & {Mamba NTP} 
% & {Mamba MLM} 
% & \begin{tabular}{@{}c@{}}Mamba MLM \\ Posthoc\end{tabular}
% & \begin{tabular}{@{}c@{}}Mamba MLM \\ RC Equiv \end{tabular} \\
% \midrule
% Mouse Enhancers & 0.716 & 0.766 & 0.745 & 0.765 & \textbf{0.790} & 0.759 \\
% Coding vs Intergenomic & 0.894 & 0.905 & 0.904 & 0.909 & \textbf{0.917} & 0.912 \\
% Human vs Worm & 0.942 & 0.964 & 0.968 & 0.970 & \textbf{0.972} & 0.969 \\
% Human Enhancer Cohn & 0.702 & 0.732 & 0.734 & 0.739 & \textbf{0.747} & 0.739 \\
% Human Enhancer Ensembl & 0.767 & 0.848 & 0.861 & 0.884 & 0.894 & \textbf{0.897} \\
% Human Regulatory & \textbf{0.872} & 0.869 & 0.822 & 0.828 & 0.708 & 0.643 \\
% Human Nontata Promoters & 0.861 & 0.942 & 0.937 & 0.935 & \textbf{0.948} & 0.947 \\
% Human OCR Ensembl & 0.698 & 0.782 & 0.805 & 0.823 & \textbf{0.834} & 0.815 \\
% \bottomrule
% \end{tabular}
% \caption{\textbf{Genomic Benchmarks} Top-1 accuracy ($\%$) for pretrained HyenaDNA, Mamba NTP, Mamba MLM, Mamba MLM Posthoc, Mamba MLM RC Equiv, and the previous SotA baseline CNN (trained from scratch).}
% \label{tab:table_vep_exp}
% \end{table*}
We begin downstream evaluation with the Genomics Benchmarks \citep{grevsova2023genomic}, a recently proposed suite with eight regulatory element classification tasks.
Non-Mamba baselines consist of HyenaDNA
% \footnote{Pre-trained weights downloaded from \url{https://huggingface.co/LongSafari/hyenadna-tiny-1k-seqlen}}
and a supervised trained CNN model described in \citet{grevsova2023genomic}
.
For HyenaDNA and all our Mamba-based models, we take the final hidden state embedding and perform mean pooling on the sequences, which vary from 200 to approximately 2,000 bps in length.
We perform 5-fold cross-validation (CV) using different random seeds, with early stopping on validation accuracy and report mean and $\pm$ on max/min of the 5 seeds.

As shown in \Tabref{tab:genomic_benchmarks}, Caduceus models attain the best performance across all annotations.
Of note, Caduceus-Ph is the best performing model overall for these tasks. 
Other works that examine post-hoc conjoining similarly find that this method attains competitive performance and often beats parameter sharing models \citep{mallet2021reverse, zhou2021towards}.

\subsubsection{Nucleotide Transformer Tasks}\label{subsubsec:nt_tasks}
% Update result chosen among the best of diff hyperparameters
\begin{table*}[ht!]
\centering
\caption{Nucleotide Transformer Tasks. Performance ($\uparrow$) across 10-fold CV for Enformer, DNABERT-2, Nucleotide Transformer v2, HyenaDNA, Caduceus-Ph, and Caduceus-PS.
Metrics vary by task: MCC for histone markers and enhancer annotation, F1-score for promoter annotation and splice site acceptor/donor, and accuracy for splice site ``all''.
Best values per task are \textbf{bolded}, second best are \textit{italicized}.
Given the disparity in model size, we also \underline{underline} the best value within the SSM-based models.
Error bars indicate the difference between the maximum and minimum values across 10 random seeds used for CV.}
\label{tab:nucleotide_transformer_task}
% \vskip 0.1in
\begin{center}
\begin{small}
\begin{sc}

\begin{tabular}{llccc||ccc}
\toprule

% {Task type} & {Dataset} 
& & \multicolumn{3}{c||}{$>$ 100M Param. Models} & \multicolumn{3}{c}{$<$ 2M Param. Models} \\
&
& \begin{tabular}{@{}c@{}}\scriptsize{Enformer} \\ \scriptsize{(252M)}\end{tabular}
& \begin{tabular}{@{}c@{}}\scriptsize{DNABERT-2} \\ \scriptsize{(117M)}\end{tabular}
% & \begin{tabular}{@{}c@{}}\small{NT-v2} \\ \small{(50M)}\end{tabular}
& \begin{tabular}{@{}c@{}}\scriptsize{NT-v2} \\ \scriptsize{(500M)}\end{tabular}
& \begin{tabular}{@{}c@{}}\scriptsize{HyenaDNA} \\ \scriptsize{(1.6M)}\end{tabular}
& \begin{tabular}{@{}c@{}}\scriptsize{Caduceus-Ph} \\ \scriptsize{(1.9M)}\end{tabular}
& \begin{tabular}{@{}c@{}}\scriptsize{Caduceus-PS} \\ \scriptsize{(1.9M)} \end{tabular} \\

\midrule
% \multirow{8}{*}{\begin{tabular}{@{}c@{}}\textit{Histone Markers}\end{tabular}}\\
\multicolumn{8}{l}{\small{\textit{Histone Markers}}} \\
 & \scriptsize{H3} & 

 \scriptsize{$0.719$}\tiny{$\pm0.048$} &
 \scriptsize{$0.785$}\tiny{$\pm0.033$} & 
 % \scriptsize{$\it{0.788}$}\tiny{$\pm0.025$} & 
 \scriptsize{$0.784$}\tiny{$\pm0.047$} &
 \scriptsize{$0.779$}\tiny{$\pm0.037$} &
 \scriptsize{\underline{$\textbf{0.815}$}}\tiny{$\pm0.048$} & \scriptsize{$\it{0.799}$}\tiny{$\pm0.029$} \\ 
& \scriptsize{H3k14ac} & \scriptsize{$0.288$}\tiny{$\pm0.077$} & \scriptsize{$0.516$}\tiny{$\pm0.028$} & 
% \scriptsize{$0.511$}\tiny{$\pm0.024$} &
\scriptsize{$0.551$}\tiny{$\pm0.021$} & \scriptsize{$\it{0.612}$}\tiny{$\pm0.065$} & \scriptsize{\underline{$\textbf{0.631}$}}\tiny{$\pm0.026$} & \scriptsize{$0.541$}\tiny{$\pm0.212$} \\ 
& \scriptsize{H3k36me3} & \scriptsize{$0.344$}\tiny{$\pm0.055$} & \scriptsize{$0.591$}\tiny{$\pm0.020$} & 
% \scriptsize{$0.583$}\tiny{$\pm0.024$} &
\scriptsize{$\textbf{0.625}$}\tiny{$\pm0.013$} & \scriptsize{\underline{$\it{0.613}$}}\tiny{$\pm0.041$} & \scriptsize{$0.601$}\tiny{$\pm0.129$} & \scriptsize{$0.609$}\tiny{$\pm0.109$} \\ 
& \scriptsize{H3k4me1} & \scriptsize{$0.291$}\tiny{$\pm0.061$} & \scriptsize{$0.511$}\tiny{$\pm0.028$} & 
% \scriptsize{$0.516$}\tiny{$\pm0.023$} & 
\scriptsize{$\textbf{0.550}$}\tiny{$\pm0.021$} & \scriptsize{$0.512$}\tiny{$\pm0.024$} & \scriptsize{\underline{$\it{0.523}$}}\tiny{$\pm0.039$} & \scriptsize{$0.488$}\tiny{$\pm0.102$} \\ 
& \scriptsize{H3k4me2} & \scriptsize{$0.211$}\tiny{$\pm0.069$} & \scriptsize{$0.336$}\tiny{$\pm0.040$} & 
% \scriptsize{$0.298$}\tiny{$\pm0.050$} &
\scriptsize{$0.319$}\tiny{$\pm0.045$} & \scriptsize{$\it{0.455}$}\tiny{$\pm0.095$} & \scriptsize{\underline{$\textbf{0.487}$}}\tiny{$\pm0.170$} & \scriptsize{$0.388$}\tiny{$\pm0.101$} \\ 
& \scriptsize{H3k4me3} & \scriptsize{$0.158$}\tiny{$\pm0.072$} & \scriptsize{$0.352$}\tiny{$\pm0.077$} & 
% \scriptsize{$0.331$}\tiny{$\pm0.049$} & 
\scriptsize{$0.410$}\tiny{$\pm0.033$} & \scriptsize\underline{{$\textbf{0.549}$}}\tiny{$\pm0.056$} & \scriptsize{$\it{0.544}$}\tiny{$\pm0.045$} & \scriptsize{$0.440$}\tiny{$\pm0.202$} \\ 
& \scriptsize{H3k79me3} & \scriptsize{$0.496$}\tiny{$\pm0.042$} & \scriptsize{$0.613$}\tiny{$\pm0.030$} & 
% \scriptsize{$0.596$}\tiny{$\pm0.022$} & 
\scriptsize{$0.626$}\tiny{$\pm0.026$} & \scriptsize{${0.672}$}\tiny{$\pm0.048$} & \scriptsize\underline{{$\textbf{0.697}$}}\tiny{$\pm0.077$} & \scriptsize{$\it{0.676}$}\tiny{$\pm0.026$} \\ 
& \scriptsize{H3K9ac} & \scriptsize{$0.420$}\tiny{$\pm0.063$} & \scriptsize{$0.542$}\tiny{$\pm0.029$} & 
% \scriptsize{$0.596$}\tiny{$\pm0.022$} & 
\scriptsize{$0.562$}\tiny{$\pm0.040$} & \scriptsize{${0.581}$}\tiny{$\pm0.061$} & \scriptsize{\underline{$\textbf{0.622}$}}\tiny{$\pm0.030$} & \scriptsize{$\it{0.604}$}\tiny{$\pm0.048$} \\ 
& \scriptsize{H4} & \scriptsize{$0.732$}\tiny{$\pm0.076$} & \scriptsize{$0.796$}\tiny{$\pm0.027$} & 
% \scriptsize{$\it{0.804}$}\tiny{$\pm0.029$} & 
\scriptsize{$\it{0.799}$}\tiny{$\pm0.025$} & \scriptsize{$0.763$}\tiny{$\pm0.044$} & \scriptsize{\underline{$\textbf{0.811}$}}\tiny{$\pm0.022$} & \scriptsize{$0.789$}\tiny{$\pm0.020$} \\ 
& \scriptsize{H4ac} & \scriptsize{$0.273$}\tiny{$\pm0.063$} & \scriptsize{$0.463$}\tiny{$\pm0.041$} & 
% \scriptsize{$0.460$}\tiny{$\pm0.039$} & 
\scriptsize{$0.495$}\tiny{$\pm0.032$} & \scriptsize{$\it{0.564}$}\tiny{$\pm0.038$} & \scriptsize{\underline{$\textbf{0.621}$}}\tiny{$\pm0.054$} & \scriptsize{$0.525$}\tiny{$\pm0.240$} \\ 

\midrule
% \multirow{5}{*}{\begin{tabular}{@{}c@{}}\small{Regulatory} \\ 
% \small{annotation}\end{tabular}}
% \multirow{5}{*}{\begin{tabular}{@{}c@{}}\small{Regulatory}\end{tabular}
\multicolumn{8}{l}{\textit{Regulatory Annotation}} \\
& \scriptsize{Enhancer} & \scriptsize{$0.451$}\tiny{$\pm0.108$} & \scriptsize{$0.516$}\tiny{$\pm0.098$} & 
% \scriptsize{$0.517$}\tiny{$\pm0.105$} & 
\scriptsize{$\textbf{0.548}$}\tiny{$\pm0.144$} & \scriptsize{$0.517$}\tiny{$\pm0.117$} & \scriptsize{\underline{$\it{0.546}$}}\tiny{$\pm0.073$} & \scriptsize{$0.491$}\tiny{$\pm0.066$} \\ 
% & \begin{tabular}{@{}c@{}}Enhancers \\ types\end{tabular}
& \scriptsize{Enhancer types} & \scriptsize{$0.309$}\tiny{$\pm0.134$} & \scriptsize{$0.423$}\tiny{$\pm0.051$} & 
% \scriptsize{$0.404$}\tiny{$\pm0.082$} & 
\scriptsize{$\it{0.424}$}\tiny{$\pm0.132$} & \scriptsize{$0.386$}\tiny{$\pm0.185$} & \scriptsize{\underline{$\textbf{0.439}$}}\tiny{$\pm0.054$} & \scriptsize{$0.416$}\tiny{$\pm0.095$} \\ 
& \scriptsize{Promoter: All} & \scriptsize{$0.954$}\tiny{$\pm0.006$} & \scriptsize{$\it{0.971}$}\tiny{$\pm0.006$} & 
% \scriptsize{$0.961$}\tiny{$\pm0.004$} & 
\scriptsize{$\textbf{0.976}$}\tiny{$\pm0.006$} & \scriptsize{$0.960$}\tiny{$\pm0.005$} & \scriptsize{\underline{$0.970$}}\tiny{$\pm0.004$} & \scriptsize{$0.967$}\tiny{$\pm0.004$} \\ 
& \scriptsize{~~~~ NonTATA} & \scriptsize{$0.955$}\tiny{$\pm0.010$} & \scriptsize{$\it{0.972}$}\tiny{$\pm0.005$} & 
% \scriptsize{$0.961$}\tiny{$\pm0.006$} & 
\scriptsize{$\textbf{0.976}$}\tiny{$\pm0.005$} & \scriptsize{$0.959$}\tiny{$\pm0.008$} & \scriptsize{\underline{$0.969$}}\tiny{$\pm0.011$} & \scriptsize{$0.968$}\tiny{$\pm0.006$} \\ 
& \scriptsize{~~~~ TATA} & \scriptsize{$\it{0.960}$}\tiny{$\pm0.023$} & \scriptsize{$0.955$}\tiny{$\pm0.021$} & 
% \scriptsize{$0.947$}\tiny{$\pm0.015$} & 
\scriptsize{$\textbf{0.966}$}\tiny{$\pm0.013$} & \scriptsize{$0.944$}\tiny{$\pm0.040$} & \scriptsize{$0.953$}\tiny{$\pm0.016$} & \scriptsize{\underline{$0.957$}}\tiny{$\pm0.015$} \\

\midrule
% \multirow{3}{*}{\begin{tabular}{@{}c@{}}\small{Splice site} \\ \small{annotation}\end{tabular}} & 
\multicolumn{8}{l}{\textit{Splice Site Annotation}} \\
& \scriptsize{All} & \scriptsize{$0.848$}\tiny{$\pm0.019$} & \scriptsize{$0.939$}\tiny{$\pm0.009$} & 
% \scriptsize{$\it{0.975}$}\tiny{$\pm0.007$} & 
\scriptsize{$\textbf{0.983}$}\tiny{$\pm0.008$} & \scriptsize{\underline{$\it{0.956}$}}\tiny{$\pm0.011$} & \scriptsize{$0.940$}\tiny{$\pm0.027$} & \scriptsize{$0.927$}\tiny{$\pm0.021$} \\
& \scriptsize{Acceptor} & 
\scriptsize{$0.914$}\tiny{$\pm0.028$} & \scriptsize{$\it{0.975}$}\tiny{$\pm0.006$} & 
% \scriptsize{$\it{0.976}$}\tiny{$\pm0.009$} & 
\scriptsize{$\textbf{0.981}$}\tiny{$\pm0.011$} & \scriptsize{\underline{$0.958$}}\tiny{$\pm0.010$} & \scriptsize{$0.937$}\tiny{$\pm0.033$} & \scriptsize{$0.936$}\tiny{$\pm0.077$} \\ 
& \scriptsize{Donor} & \scriptsize{$0.906$}\tiny{$\pm0.027$} & \scriptsize{$\it{0.963}$}\tiny{$\pm0.006$} & 
% \scriptsize{$\it{0.973}$}\tiny{$\pm0.010$} & 
\scriptsize{$\textbf{0.985}$}\tiny{$\pm0.022$} & \scriptsize{\underline{$0.949$}}\tiny{$\pm0.024$} & \scriptsize{$0.948$}\tiny{$\pm0.025$} & \scriptsize{$0.874$}\tiny{$\pm0.289$} \\  
\bottomrule
\end{tabular}
\end{sc}
\end{small}
\end{center}
\vskip -0.1in
\end{table*}

Next, we benchmark against a collection of 18 datasets introduced in \citet{dalla2023nucleotide} and derived from five peer-reviewed studies \citep{phaml2005qualitatively, oubounyt2019deepromoter, wang2019splicefinder, scalzitti2021spliceator, geng2022deep}.
These datasets contain three task types, including histone marker prediction, regulatory annotation prediction, and splice site annotation prediction.
In assessing performance, we adhered to the methodology described in \citet{dalla2023nucleotide}, using different metrics for the tasks: Matthews Correlation Coefficient (MCC) for all histone marker tasks and enhancer classification, F1 score for promoter regulatory annotations, and splice site annotation tasks, except for the \textit{splice sites all} task, where we report accuracy.
We additionally follow \citet{dalla2023nucleotide} in performing 10-fold CV using different random seeds with early stopping on the validation metric.
We report mean and $\pm$ on max/min of the 10 seeds.
% \footnote{Baseline models' performance are taken from \url{https://huggingface.co/spaces/InstaDeepAI/nucleotide_transformer_benchmark}}.
The results for this benchmark suite are presented in \Tabref{tab:nucleotide_transformer_task}, where we again find that Caduceus-Ph performs competitively, even beating attention-based methods with orders of magnitude more parameters on 8 of 18 prediction tasks.
Caduceus models outperform a similarly sized HyenaDNA 
model on almost all the histone marker and regulatory annotation tasks, while HyenaDNA performs better on splice site annotation.

\subsubsection{Predicting the Effect of Variants on Gene Expression}\label{subsubsec:vep}
Finally, we explore the implications of long-range contexts on the task of predicting the effect of SNPs on gene expression.
There is biological evidence to suggest this task indeed entails long-range interactions \citep{furlong2018_enhancers}.
Additionally it aligns well to LM pre-training objectives, which enable models to implicitly learn to recognize the effects of evolutionary pressure (e.g., conservation, co-evolution).
The dataset used in this task is derived from the Enformer paper \citep{avsec2021effective} and presented in \citet{trop2023LRB}.
From each model, we extract embeddings centered around the SNP location.
We stratify the data by distance of the SNP to nearest Transcription Start Site (TSS).
For each bucket, we sample 5,000 training points and fit an SVM classifier with an RBF kernel to predict VEP annotations.
We report test set AUCROC mean $\pm$ standard deviation for classifiers fit on 5 random training subsets.
For more details about this experiment, please refer to \Appsecref{appsubsec:vep}.
We compare Caduceus to HyenaDNA
% \footnote{Pre-trained weights downloaded from \url{https://huggingface.co/LongSafari/hyenadna-medium-160k-seqlen-hf}.}
and Nucleotide Transformer,
% \footnote{Pre-trained weights downloaded from \url{https://huggingface.co/InstaDeepAI/nucleotide-transformer-v2-500m-multi-species}.}
as well as to the supervised baseline Enformer \citep{avsec2021effective}.
% \footnote{Pre-trained weights downloaded from \url{https://huggingface.co/EleutherAI/enformer-official-rough}.} 

\begin{figure*}[ht!]
    \centering
    \includegraphics[width=0.9\linewidth]{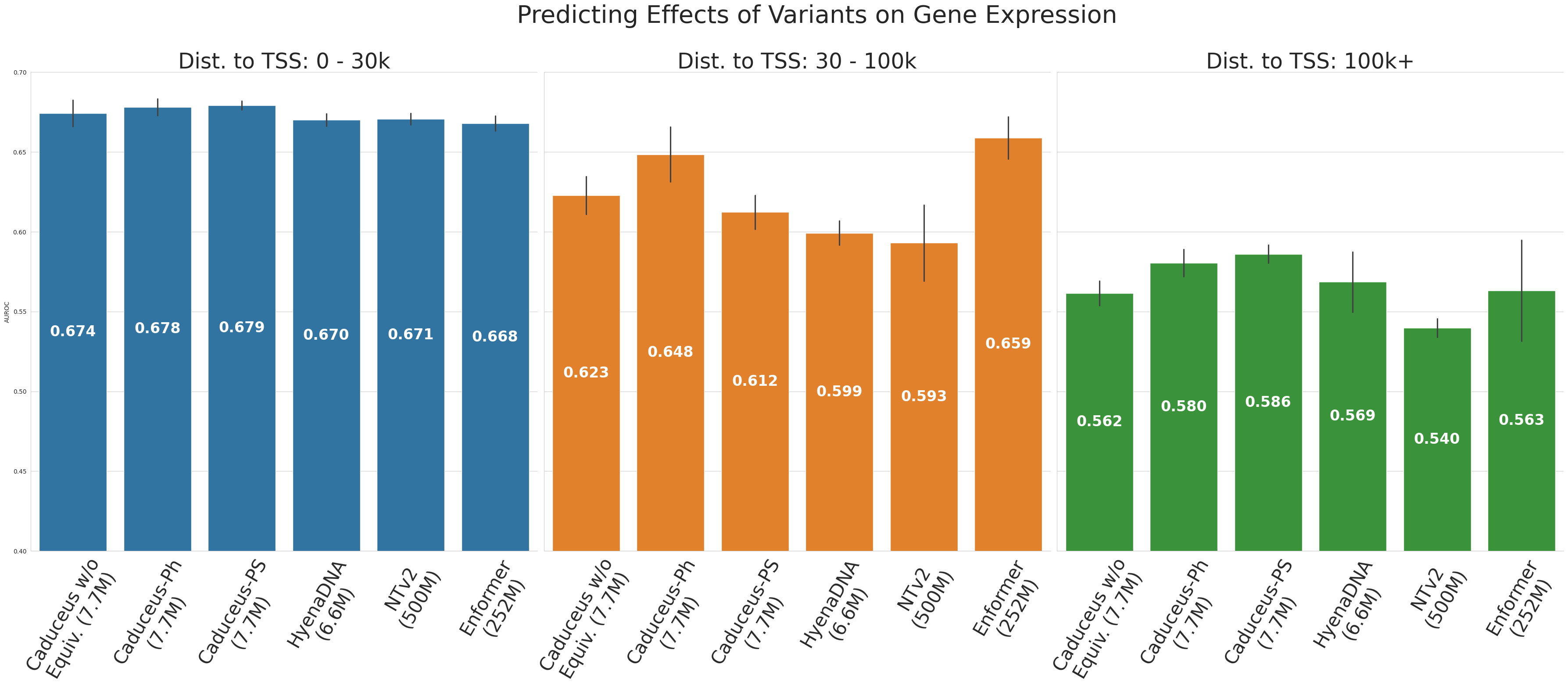}
    \caption{Predicting variant effects on gene expression across varying distances to the nearest Transcription Start Site (TSS). Models compared include Enformer, NT-v2, HyenaDNA, Caduceus w/o RC Equiv, Caduceus-Ph, and Caduceus-PS, with model sizes indicated in parentheses.
    SSM-based models utilize a 131k sequence length.
    We show performance at short (0 - 30k bps), medium (30 - 100k bps), and long-range ($>$100k bps) distances to TSS. Notably, Caduceus-PS consistently demonstrates enhanced predictive accuracy for long-range effects.
    Error bars represent standard deviation across five SVM classifiers, each fit on different dataset subsets.}
    \label{fig:variant_effect_prediction}
\end{figure*}

As shown in \Figref{fig:variant_effect_prediction}, Caduceus models consistently outperform HyenaDNA, and Caduceus-PS exceeds the performance of Nucleotide Transformer v2 (with 500M parameters), especially as distance to the nearest TSS grows.
Of note, on sequences where distance to TSS exceeds 100k, Caduceus even outperforms the well-regarded Enformer baseline.

\section{Related Work}\label{sec:related}

\subsection{DNA Language Models}

% \paragraph{Transformer-based}
% A significant challenge in deploying Transformer models in genomics is dealing with the long context of DNA sequences due to the quadratic nature of the multi-head attention mechanism. While a meaningful gene sequence may span several kilobases, effectively employing Transformers requires specific design considerations within the model.
Transformer-based DNA LMs, such as DNABERT, v1 \citep{ji2021dnabert} and v2 \citep{zhou2023dnabert}, and Nucleotide Transformer \citep{dalla2023nucleotide} have been restricted by the quadratic scaling of Transformers, with maximum context sizes of up to roughly 12,000 bps.
BigBird \citep{zaheer2020big} (and GENA-LM \citep{fishman2023gena}, which uses BigBird as a backbone) use sparse attention to scale context size up to an order of magnitude large.

Notably, GPN \citep{benegas2023gpnmsa,benegas2023gpn}, uses dilated convolutional layers, which in practice scale to large receptive fields, although a context size of only 512 bps is used when training this model.
\citet{benegas2023gpn} find that DNA LMs are powerful unsupervised variant effect predictors.

\paragraph{HyenaDNA}
Most related to our work is the HyenaDNA model \citep{nguyen2023hyenadna}, which uses the Hyena operator \citep{poli2023hyena}, derived from the SSM literature, as the building block for a DNA LM.
HyenaDNA is able to scale to long-range sequences (up to 1 million bps), but is uni-directional and not inherently robust to RC inputs. 

\subsection{Reverse Complement Training for DNA}
\citet{cao2019simple} discuss the importance of RC data augmentation in genomics.
\citet{shrikumar2017reverse} introduce RC Parameter Sharing (RCPS) for convolution, batch normalization, and pooling modules.
% Notably, in \citet{shrikumar2017reverse}, the RCPS layers effectively double the channel dimension after each layer.
\citet{mallet2021reverse} formalize RC equivariance in the language of Group representations and cast RCPS as a particular decomposition of such representations, exploring other as well.
Our implementation of RCPS in the MambaDNA block differs from that proposed in \citet{shrikumar2017reverse} in that our split operation prevents the channel dimension from doubling when passing a sequence through a given layer.

\citet{zhou2021towards} further explore RCPS layers and compare them to a \textit{post-hoc} conjoining baseline, which serves as the inspiration for our Caduceus-Ph model.
\citet{zhou2021towards} find that post-hoc conjoining is a strong baseline that often outperforms RCPS models on several tasks.
We note that \citet{zhou2021towards} focus on supervised training regimes, whereas we extend the post-hoc conjoining methodology to include a LM pre-training step as well.
Prediction conjoining was also explored in DeepBind \citep{alipanahi2015predicting}, where max aggregation as opposed to averaging is used, and in FactorNet \citep{quang2019factornet}, which performs conjoining during training and inference.

Finally, \citet{gunduz2023self} also explore RC sequences in self-supervised pre-training.
However, their model uses contrastive learning where an encoder is trained to recognize the embeddings of the RC sequence in a given batch.

% To better harness the symmetry of DNA, researchers have developed specialized models that explicitly account for the RC characteristics. Two notable approaches are the \textit{conjoined model}~\citep{alipanahi2015predicting, quang2019factornet} and the \textit{Reverse Complement Parameter Sharin}g (RCPS) model~\citep{shrikumar2017reverse, mallet2021reverse, zhou2021towards}. The conjoined model processes both the forward and reverse complement sequences through an identical shared, i.e., siamese, model, averaging the predictions from both branches before the final layer. On the other hand, the RCPS model pairs each "forward" filter with a corresponding "RC" filter. These paired filters are designed to recognize patterns and their reverse complements, effectively learning to identify motifs irrespective of their orientation on the DNA strand. Despite the fact that both the post-hoc conjoined model and RCPS model take into account the RC equivalence, it is worth noting that it has been shown that the post-hoc conjoined model performs comparably or even better than the RCPS model, a difference that is not easily attributed to overfitting~\citep{bartoszewicz2020deepac, zhou2021towards}.

\subsection{Bi-directional RNNs}
Exploiting bi-directionality for pre-training on large datasets was first realized in ELMo \citep{peters2017semi}, where forward and backward LSTMs \citep{hochreiter1997long} were utilized simultaneously to model language context.
This laid the groundwork for models such as BERT \citep{devlin2018bert} that replaced recurrent networks with a Transformer backbone.
Recently, \citet{wang2022pretraining} explored BERT-style training using SSMs.
In concurrent work, \citet{zhu2024vision} also extend the Mamba SSM to be bi-directional, similarly combining outputs of forward and backward sequence operators.

\section{Conclusion}\label{sec:conclusion}
In this work, we introduced architectural innovations to the Mamba module, enabling bi-directional and RC equivariant sequence modeling.
We also propose a new DNA foundation model, Caduceus, and demonstrate its ability to outperform comparably sized uni-directional Hyena-based models and Transformer-based models orders of magnitude larger in size on a range of biologically relevant tasks, most notably predicting the effect of genetic mutations on gene expression.

\section*{Impact Statement}
This paper presents work whose goal is to advance the field of machine learning.
There are many potential societal consequences of our work.
As with all machine learning models, and particularly language models, our work has the potential for societal benefits but can be subject to misuse.

\section*{Acknowledgments}
This work was supported by an NSF CAREER grant (\#2145577) and an NIH MIRA grant (\#1R35GM151243-01).
We would also like to thank Evan Trop and the InstaDeep team for useful discussions about the Nucleotide Transformer leaderboard and the variant effect prediction task and MosaicML for providing compute resources for some of the pre-training experiments.

\bibliography{refs}
\bibliographystyle{icml2024}

%%%%%%%%%%%%%%%%%%%%%%%%%%%%%%%%%%%%%%%%%%%%%%%%%%%%%%%%%%%%%%%%%%%%%%%%%%%%%%%
%%%%%%%%%%%%%%%%%%%%%%%%%%%%%%%%%%%%%%%%%%%%%%%%%%%%%%%%%%%%%%%%%%%%%%%%%%%%%%%
% APPENDIX
%%%%%%%%%%%%%%%%%%%%%%%%%%%%%%%%%%%%%%%%%%%%%%%%%%%%%%%%%%%%%%%%%%%%%%%%%%%%%%%
%%%%%%%%%%%%%%%%%%%%%%%%%%%%%%%%%%%%%%%%%%%%%%%%%%%%%%%%%%%%%%%%%%%%%%%%%%%%%%%
\newpage
\appendix
\onecolumn
\section{Proof of \Thrmref{thrm:mamba_rcequiv}}\label{appsec:mamba_rcequiv_proof}
We begin by reiterating the definitions of the different functions that comprise our RC equivariant Mamba module.
For an input sequence $\rmX_{1:T}^{1:D}$ of length $T$, with $D$ channels, we define:
\begin{align}
    \splitchan(\rmX_{1:T}^{1:D}) &:= \left[\rmX_{1:T}^{1:(D/2)}, \rmX_{1:T}^{(D/2):D}\right], \label{def:split} \\
    \RC\left(\rmX_{1:T}^{1:D}\right) &:= \rmX_{T:1}^{D:1}, \label{def:rc} \\
    \concat\left(\left[
    \rmX_{1:T}^{1:(D/2)}, \rmX_{1:T}^{(D/2):D}\right]\right) &:= \rmX_{1:T}^{1:D} \label{def:concat}, \\
    \M_{\RCe,\theta}\left(\rmX_{1:T}^{1:D}\right) &:= \concat\left(\left[\M_\theta\left(\rmX_{1:T}^{1:(D/2)}\right), \RC\left(\M_\theta\circ\RC\left(\rmX_{1:T}^{(D/2):D}\right)\right)\right]\right) \label{def:mamba_rc}.
\end{align}
We also denote
% the application of a Mamba module $\M_\theta$ to a sequence that is `split' along the channel dimension as follows:
% \begin{align}
%     \M_\theta\left(\left[
%     \rmX_{1:T}^{1:(D/2)}, \rmX_{1:T}^{(D/2):D}\right]\right) &:= \left[\M_\theta\left(\rmX_{1:T}^{1:(D/2)}\right), \M_\theta\left(\rmX_{1:T}^{(D/2):D}\right)\right], \label{def:mamba_split}
% \end{align}
% and
the application of the $\RC$ operation to a sequence that is `split' along the channel dimension as:
\begin{align}
    \RC\left(\left[
    \rmX_{1:T}^{1:(D/2)}, \rmX_{1:T}^{(D/2):D}\right]\right) &:= \left[\RC\left(\rmX_{1:T}^{(D/2):D}\right), \RC\left(\rmX_{1:T}^{(1:(D/2)}\right)\right] = \left[\rmX_{T:1}^{D:(D/2)}, \rmX_{T:1}^{(D/2):1}\right], \label{def:rc_split}
\end{align}

% Note that the $\RC$ operation commutes with the $\splitchan$ operation, since:
% \begin{align}
% \begin{split}\label{lemma:rc_split} 
%     \RC\circ\splitchan\left(
%     \rmX_{1:T}^{1:D}\right) &= \left[\rmX_{T:1}^{D:(D/2)}, \rmX_{T:1}^{(D/2):1}\right] \\
%     &= \splitchan\left(
%     \rmX_{T:1}^{D:1}\right) = \splitchan\circ\RC\left(
%     \rmX_{1:T}^{1:D}\right).
% \end{split}
% \end{align}
% Additionally, 
Note that the $\RC$ operation can be `pulled inside' of a $\concat$ operation:
\begin{align}
\begin{split}\label{lemma:rc_concat}
    \RC\circ\concat\left(\left[\rmX_{1:T}^{1:(D/2)}, \rmX_{1:T}^{(D/2):D}\right]\right)
    &= \RC\left(\rmX_{1:T}^{1:D}\right) \\
    &= \rmX_{T:1}^{D:1} \\
    &= \concat\left(\left[\rmX_{T:1}^{D:(D/2)}, \rmX_{T:1}^{(D/2):1}\right]\right) \\
    &= \concat\left(\left[\RC\left(\rmX_{1:T}^{1:(D/2)}\right), \RC\left(\rmX_{1:T}^{(D/2):D}\right)\right]\right) \\
    &= \concat\circ\RC\left(\left[\rmX_{1:T}^{1:(D/2)}, \rmX_{1:T}^{(D/2):D}\right]\right)
\end{split}
\end{align}
Additionally, we have that $\RC^{-1} = \RC$ and that
\begin{align}\label{lemma:rc_undo}
    \RC\left(\left[\rmX_{1:T}^{1:(D/2)}, \RC\left(\rmX_{1:T}^{(D/2):D} \right) \right]\right) = \left[\rmX_{1:T}^{(D/2):D}, \RC\left(\rmX_{1:T}^{1:(D/2)} \right) \right].
\end{align}

Following Definition \ref{def:mamba_rc}, we have that:
\begin{align*}
    \RC\circ\M_{\RCe, \theta}\left(\rmX_{1:T}^{1:D}\right) &= \RC\circ\concat\left(\left[\M_\theta\left(\rmX_{1:T}^{1:(D/2)}\right), \RC\left(\M_\theta\circ\RC\left(\rmX_{1:T}^{(D/2):D}\right)\right)\right]\right) && \textit{(\ref{def:mamba_rc})}\\
    &= \concat\circ\RC\left(\left[M_\theta\left(\rmX_{1:T}^{1:(D/2)}\right), \RC\left(\M_\theta\circ\RC\left(\rmX_{1:T}^{(D/2):D}\right)\right)\right]\right) && \textit{(\ref{lemma:rc_concat})} \\
    &= \concat\left(\left[\M_\theta\circ\RC\left(\rmX_{1:T}^{(D/2):D}\right), \RC\left(\M_\theta\left(\rmX_{1:T}^{1:(D/2)}\right)\right)\right]\right) && \textit{(\ref{lemma:rc_undo})} \\
    &= \concat\left(\left[\M_\theta\left(\rmX_{T:1}^{D:(D/2)}\right), \RC\left(\M_\theta\circ\RC\left(\rmX_{T:1}^{(D/2):1}\right)\right)\right]\right) && \textit{(\ref{def:rc})} \\
    &= \M_{\RCe,\theta}\circ\RC\left(\rmX_{1:T}^{1:D}\right) && \square
\end{align*}

\section{Proof of \Thrmref{thrm:caduceus_rcequiv}}\label{appsec:caduceus_rcequiv_proof}
We begin with the following lemma,
\begin{lemma}\label{lemma:rc_compose}
    For two RC equivariant sequence operators $\mathrm{F}$ and $\mathrm{G}$, their composition $\mathrm{F} \circ \mathrm{G}$ is also equivariant.
\end{lemma}
\textit{Proof.} We have that,
\begin{align*}
    \mathrm{F}\left(\mathrm{G}\left(\RC\left(\rmX_{1:T}^{1:D}\right)\right)\right) = \mathrm{F}\left(\RC\left(\mathrm{G}\left(\rmX_{1:T}^{1:D}\right)\right)\right) = \RC\left(\mathrm{F}\left(\mathrm{G}\left(\rmX_{1:T}^{1:D}\right)\right)\right)
\end{align*}
where each equality follows from the RC equivariance of the operators $\mathrm{G}$ and $\mathrm{F}$, respectively. \hfill $\square$

Therefore, to prove that the Caduceus-PS is RC equivariant, we need to prove that each operator in  $\LM_{\RCe, \theta} \circ \M_{\RCe, \theta}^{(n)} \circ \Emb_{\RCe, \theta}$ satisfies this property.

First, we show that  $\Emb_{\RCe, \theta}$ is RC equivariant.
\begin{align}\label{lemma:emb_rcequiv}
    \RC \circ \Emb_{\RCe, \theta}\left(\rmX_{1:T}^{1:4}\right) &= 
    \RC\circ\concat\left(\left[\Emb_\theta\left(\rmX_{1:T}^{1:4}\right), \RC\circ\Emb_\theta\left(\RC\left(\rmX_{1:T}^{1:4}\right)\right)\right]\right) \nonumber \\
    &= \concat\circ\RC\left(\left[\Emb_\theta\left(\rmX_{1:T}^{1:4}\right), \RC\circ\Emb_\theta\left(\RC\left(\rmX_{1:T}^{1:4}\right)\right)\right]\right) && \textit{(\ref{lemma:rc_concat})} \nonumber \\
    &= \concat\left(\left[\Emb_\theta\left(\RC\left(\rmX_{1:T}^{1:4}\right)\right), \RC\circ\Emb_\theta\left(\rmX_{1:T}^{1:4}\right)\right)\right] && \textit{(\ref{lemma:rc_undo})} \nonumber \\
    &= \concat\left(\left[\Emb_\theta\left(\rmX_{T:1}^{4:1}\right), \RC\circ\Emb_\theta\left(\rmX_{1:T}^{1:4}\right)\right)\right] \nonumber \\
    &= \concat\left(\left[\Emb_\theta\left(\rmX_{T:1}^{4:1}\right), \RC\circ\Emb_\theta\left(\RC\left(\rmX_{T:1}^{4:1}\right)\right)\right)\right] \nonumber \\
    &= \Emb_{\RCe,\theta} \circ \RC\left(\rmX_{1:T}^{1:4}\right) &&\square
\end{align}

Additionally, we have that $\M_{\RCe, \theta}^{(n)}$ is equivariant by \Thrmref{thrm:mamba_rcequiv} and induction using Lemma \ref{lemma:rc_compose}.

Finally, recall the definition of $\LM_{\RCe, \theta}$:
\begin{align*}
    \LM_{\RCe, \theta}\left(\rmX_{1:T}^{1:D}\right) :=
    \LM_\theta\left(\rmX_{1:T}^{1:(D/2)}\right) + \flipchan\circ\LM_\theta\left(\rmX_{1:T}^{D:(D/2)}\right).
\end{align*}
Note that $\LM_\theta$ is parameterized by a weight matrix $\mW_\theta$ and applying $\LM_\theta$ to a sequence $\rmX_{1:T}^{1:(D/2)}$ is equivalent to multiplying each of the sequence elements $\rvx_t^{1:(D/2)},$ for $t = 1, \ldots, T,$ on the left by $\mW_\theta.$
Therefore if we reverse an input to $\LM_\theta$ along the length dimension, the output will be reversed along the length dimension as well.
We can thus focus on a specific item at position $t$ in a sequence:
\begin{align*}
    \LM_{\RCe, \theta}\left(\rmX_{1:T}^{1:D}\right)_t = \mW_\theta\cdot\rvx_t^{1:(D/2)} + \flipchan\left(\mW_\theta\cdot\rvx_t^{D:(D/2)}\right),
\end{align*}
and we need only show that it is equivariant with the $\flipchan$ operation, which we recall merely reverses the channels of given input.
We note that $\flipchan^{-1} = \flipchan.$
Now we show that:
\begin{align*}\label{lemma:lm_rcequiv}
\begin{split}
    \flipchan\left(\LM_{\RCe, \theta}\left(\rmX_{1:T}^{1:D}\right)_t\right) &= \flipchan\left(\mW_\theta\cdot\rvx_t^{1:(D/2)}\right) + \mW_\theta\cdot\rvx_t^{D:(D/2)} \\
    &= \LM_{\RCe, \theta}\left(\flipchan\left(\rmX_{1:T}^{1:D}\right)\right)_t
\end{split}
\end{align*}
This completes the proof. \hfill $\square$

\section{Pre-training}\label{appsec:pretrain}
We provide a more detailed description of the dataset and training methodology used in the human reference genome pre-training task.
This dataset is based on the splits used in the previous Enformer study \citep{avsec2021effective}.
The training split comprises 34,021 segments that we extend to a maximum length of 1,048,576 ($2^{20}$), collectively covering the genome and amounting to around 35 billion tokens, or nucleotide base pairs.
% These segments are identified by a trio of values: chromosome number, start index, and end index, and can be lengthened if needed to obtain longer segments.

All the Mamba-based models, including Caduceus, were trained with a learning rate of $8\mathrm{e}^{-3}$.
We maintain a constant number of tokens in each batch, using $2^{20}$ tokens per batch.
For example, for sequence lengths of 1,024, batch size is also 1,024 and for sequence lengths of 131k ($2^{17}$), batch size is 8.
All our models, other than Caduceus-PS, are pre-trained with RC data augmentation, where any given sequence is either unchanged or has the RC operation applied to it with equal probability.

Models were trained with cosine decay and the \textsc{ADAM} optimization algorithm \citep{kingma2014adam}, $\beta_1$ and $\beta_2$ values of 0.95 and 0.9, respectively.

For bi-directional models, we use the masking recipe presented in \citet{devlin2018bert}.
Namely, we `mask' 15\% of tokens.
Of the `masked' tokens, 80\% are replaced with a special \textsc{[MASK]} token, 10\% are replaced with a random token from the vocabulary, and 10\% are left unchanged.

The various Mamba/Caduceus models that were pre-trained are listed in 
\Tabref{tab:pretrained}.
For Figure \ref{subfig:mamba_ntp_vs_hyenadna}, we re-pre-train HyenaDNA models on sequence lengths of 1,024, 32k, and 131k.
We use the corresponding hidden dimension and depth as those used when these models were originally trained in \citet{nguyen2023hyenadna}.
Other than learning rate, which was set to $6\mathrm{e}^{-4}$, all the other pre-training details used for our models above were used for HyenaDNA pre-training as well.

\begin{table}[ht]
\caption{Pre-trained Mamba-based models with corresponding sequence length, depth, hidden dimension, and number of gradient updates.}
\begin{center}
\begin{small}
\begin{sc}
\begin{tabular}{ccccccc}
\toprule
Seq. Len. & Hidden Dim. & Num. Layers & Gradient Updates & Uni & Bi-directional &\begin{tabular}{@{}c@{}}\small{Bi-directional} \\ \small{RC Equiv.} \end{tabular}  \\
\midrule
1k & 118 & 4 & 10k & & $\checkmark$ & $\checkmark$ \\
1k & 128 & 4 & 10k & $\checkmark$ & $\checkmark$ & $\checkmark$ \\
1k & 256 & 4 & 10k & $\checkmark$ & $\checkmark$ & $\checkmark$ \\
32k & 256 & 8 & 10k & $\checkmark$ & $\checkmark$ & $\checkmark$ \\
131k & 256 & 16 & 50k & $\checkmark$ & $\checkmark$ & $\checkmark$\\
% 131k & 384 & 20 & 50k & $\checkmark$ & $\checkmark$ & \\
\bottomrule
\end{tabular}
\end{sc}
\end{small}
\end{center}
\label{tab:pretrained}
\end{table}

\section{Downstream Tasks}\label{appsec:downstream}
\subsection{Genomics Benchmark}\label{appsubsec:gb}
For the Genomics Benchmark tasks, we deviate from the results presented in \citet{nguyen2023hyenadna} in order to maintain `true' train and test splits.
We therefore, elect to use 5-fold cross-validation where we split the training set into 90/10 train/validation splits and perform early stopping on the validation set.
Models were fine-tuned for 10 epochs.
The HyenaDNA model consists of 2 layers and hidden dimension 128.
It is fine-tuned with a learning rate of $6\mathrm{e}^{-4}$ and batch size of 256.
Weights for this pre-trained model were downloaded from \url{https://huggingface.co/LongSafari/hyenadna-tiny-1k-seqlen}.
Following \citet{nguyen2023hyenadna}, we also experiment with adding RC data augmentation for HyenaDNA.
The best result of this search is presented in \Tabref{tab:genomic_benchmarks}.
The values used for RC data augmentation in each task are presented in
\Tabref{tab:genomic_benchmarks_hyperparams_for_hyena}.

The CNN baseline is described in \citet{grevsova2023genomic}.
It is trained from scratch with a learning rate of $1\mathrm{e}^{-3}$ and batch size of 64.
The CNN consists of an embedding layer and convolutional layers with 16, 8, and 4 channels.
The first layer is followed by a ReLU non-linearity and all layers are followed by batch normalization and 1D max-pooling.
Finally there are two fully connected layers at the end of the network.

The Caduceus and Mamba models were fine-tuned with a batch size of 256.  For the learning rate, we performed hyperparameter tuning, searching within $\{1\mathrm{e}^{-3},2\mathrm{e}^{-3}\}$, and present the best result across cross-valildation, as shown in  \Tabref{tab:genomic_benchmarks_hyperparams_for_caduceus}.
Mamba models consist of 4 layers with hidden dimension 128 and Caduceus models consist of 4 layers with hidden dimension 118 (to keep parameter counts roughly equivalent).
For both Caduceus-Ph and Caduceus-PS the forward and RC sequence representations are pooled and then averaged.
For Caduceus-PS, this averaging is done during both downstream training and inference.
For Caduceus-Ph, this is done only during inference.

\begin{table}[th]
\caption{Hyena Hyperparameter Selection for Genomic Benchmarks. The HyenaDNA model, chosen for its top-1 accuracy averaged over 5-fold cross-validation, includes options for using or not using the RC data augmentation during pre-training.}
\begin{center}
\begin{small}
\begin{sc}
\begin{tabular}{lc}
\toprule
Mouse Enhancers & No RC Augmentation \\
Coding vs. Intergenomic & No RC Augmentation \\
Human vs. Worm & RC Augmentation \\
Human Enhancers Cohn & RC Augmentation \\
Human Enhancer Ensembl & No RC Augmentation \\
Human Regulatory & RC Augmentation \\
Human OCR Ensembl & RC Augmentation \\
Human NonTATA Promoters & No RC Augmentation \\
% &  \\
% \midrule
% \small{Mouse Enhancers} & No RC Augmentation \\
% \small{Coding vs Intergenomic} & RC Augmentation \\
% \small{Human vs Worm} & No RC Augmentation \\
% \small{Human Enhancer Cohn} & RC Augmentation \\
% \small{Human Enhancer Ensembl} & No RC Augmentation \\
% \small{Human Regulatory} & RC Augmentation \\
% \small{Human Nontata Promoters} & No RC Augmentation \\
% \small{Human OCR Ensembl} & RC Augmentation \\
\bottomrule
\end{tabular}
\end{sc}
\end{small}
\end{center}
\label{tab:genomic_benchmarks_hyperparams_for_hyena}
\end{table}

\begin{table}[th]
\caption{Mamba / Caduceus Hyperparameter Selection for Genomic Benchmarks. Learning rate chosen for its top-1 accuracy averaged over 5-fold cross-validation.}
\begin{center}
\begin{small}
\begin{sc}
\begin{tabular}{lcccc}
\toprule
& Mamba & Caduceus w/o Equiv. & Caduceus-Ph & Caduceus-PS \\
\midrule
Mouse Enhancers & $2\mathrm{e}^{-3}$ & $2\mathrm{e}^{-3}$ & $2\mathrm{e}^{-3}$ & $2\mathrm{e}^{-3}$ \\
Coding vs. Intergenomic & $2\mathrm{e}^{-3}$ & $1\mathrm{e}^{-3}$ & $2\mathrm{e}^{-3}$ & $1\mathrm{e}^{-3}$ \\
Human vs. Worm & $2\mathrm{e}^{-3}$ & $1\mathrm{e}^{-3}$ & $2\mathrm{e}^{-3}$ & $1\mathrm{e}^{-3}$ \\
Human Enhancers Cohn & $1\mathrm{e}^{-3}$ & $1\mathrm{e}^{-3}$ & $1\mathrm{e}^{-3}$ & $2\mathrm{e}^{-3}$ \\
Human Enhancer Ensembl & $2\mathrm{e}^{-3}$ & $1\mathrm{e}^{-3}$ & $1\mathrm{e}^{-3}$ & $1\mathrm{e}^{-3}$ \\
Human Regulatory & $1\mathrm{e}^{-3}$ & $2\mathrm{e}^{-3}$ & $2\mathrm{e}^{-3}$ & $1\mathrm{e}^{-3}$ \\
Human OCR Ensembl & $2\mathrm{e}^{-3}$ & $2\mathrm{e}^{-3}$ & $2\mathrm{e}^{-3}$ & $2\mathrm{e}^{-3}$ \\
Human NonTATA Promoters & $1\mathrm{e}^{-3}$ & $2\mathrm{e}^{-3}$ & $2\mathrm{e}^{-3}$ & $2\mathrm{e}^{-3}$ \\
% \small{Mouse Enhancers} & $2\mathrm{e}^{-3}$ & $2\mathrm{e}^{-3}$ \\
% \small{Coding vs Intergenomic} & $2\mathrm{e}^{-3}$ & $2\mathrm{e}^{-3}$ \\
% \small{Human vs Worm} & $2\mathrm{e}^{-3}$ & $2\mathrm{e}^{-3}$ \\
% \small{Human Enhancer Cohn} & $2\mathrm{e}^{-3}$ & $2\mathrm{e}^{-3}$ \\
% \small{Human Enhancer Ensembl} & $2\mathrm{e}^{-3}$ & $2\mathrm{e}^{-3}$ \\
% \small{Human Regulatory} & $1\mathrm{e}^{-3}$ & $2\mathrm{e}^{-3}$ \\
% \small{Human Nontata Promoters} & $2\mathrm{e}^{-3}$ & $2\mathrm{e}^{-3}$ \\
% \small{Human OCR Ensembl} & $2\mathrm{e}^{-3}$ & $2\mathrm{e}^{-3}$ \\
\bottomrule
\end{tabular}
\end{sc}
\end{small}
\end{center}
\label{tab:genomic_benchmarks_hyperparams_for_caduceus}
\end{table}

\subsection{Nucleotide Transformer Tasks}\label{appsubsec:nt}
For the Nucleotide Transformer Task, we pull baseline results from \url{https://huggingface.co/spaces/InstaDeepAI/nucleotide_transformer_benchmark}.
For our Caduceus / Mamba-based models we follow the same CV protocol from \citet{dalla2023nucleotide} using a 90/10 train/validation split for each fold.
Our models consist of 4 layers and hidden dimension 256, roughly matching the parameter count of the reported HyenaDNA model.
Models were fine-tuned for 20 epochs.
Hyperparameters for the models reported in \Tabref{tab:nucleotide_transformer_task} can be found in \Tabref{tab:nucleotide_transformer_task_hyperparam_for_mamba}

\begin{table*}[th]
\caption{Caduceus Hyperparameter Selection for Nucleotide Transformer Tasks. Caduceus-Ph and Caduceus-PS fine-tuning hyperparameters chosen based on best performance averaged over 10-fold cross-validation.}
\begin{center}
\begin{small}
\begin{sc}
\label{tab:nucleotide_transformer_task_hyperparam_for_mamba}
\begin{tabular}{lccccccc}
\toprule
& & \multicolumn{2}{c}{Caduceus-Ph} & \multicolumn{2}{c}{Caduceus-PS} \\
&
& LR & batch size & LR & batch size \\
\midrule
\multirow{10}{*}{\begin{tabular}{@{}c@{}}\small{Histone} \\ \small{markers}\end{tabular}}
 & \small{H3} & $1\mathrm{e}^{-3}$ & 128	& $1\mathrm{e}^{-3}$	& 128 \\ 
& \small{H3k14ac} & $1\mathrm{e}^{-3}$ & 128	& $1\mathrm{e}^{-3}$	& 128 \\ 
& \small{H3k36me3} & $1\mathrm{e}^{-3}$ & 128	& $1\mathrm{e}^{-3}$	& 128 \\ 
& \small{H3k4me1} & $1\mathrm{e}^{-3}$ & 512	& $1\mathrm{e}^{-3}$	& 128 \\ 
& \small{H3k4me2} & $1\mathrm{e}^{-3}$ & 128	& $1\mathrm{e}^{-3}$	& 512 \\ 
& \small{H3k4me3} & $1\mathrm{e}^{-3}$ & 512	& $1\mathrm{e}^{-3}$	& 512 \\ 
& \small{H3k79me3} & $1\mathrm{e}^{-3}$ & 128	& $1\mathrm{e}^{-3}$	& 128 \\ 
& \small{H3K9ac} & $1\mathrm{e}^{-3}$ & 128	& $1\mathrm{e}^{-3}$	& 128 \\ 
& \small{H4} & $1\mathrm{e}^{-3}$ & 128	& $1\mathrm{e}^{-3}$	& 128 \\ 
& \small{H4ac} & $1\mathrm{e}^{-3}$ & 128	& $1\mathrm{e}^{-3}$	& 128 \\ 

\midrule
\multirow{5}{*}{\begin{tabular}{@{}c@{}}\small{Regulatory} \\ \small{annotation}\end{tabular}}
& \small{Enhancers} & $1\mathrm{e}^{-3}$ & 512	& $1\mathrm{e}^{-3}$	& 512 \\ 
& \small{Enhancers types} & $1\mathrm{e}^{-3}$ & 512	& $2\mathrm{e}^{-3}$	& 512 \\ 
& \small{Promoter all} & $1\mathrm{e}^{-3}$ & 512	& $1\mathrm{e}^{-3}$	& 128 \\ 
& \small{Promoter no tata} & $1\mathrm{e}^{-3}$ & 512	& $1\mathrm{e}^{-3}$	& 128 \\  
& \small{Promoter tata} & $1\mathrm{e}^{-3}$ & 128	& $1\mathrm{e}^{-3}$	& 512 \\ 

\midrule
\multirow{3}{*}{\begin{tabular}{@{}c@{}}\small{Splice site} \\ \small{annotation}\end{tabular}} & 
\small{Splice sites acceptors} & $1\mathrm{e}^{-3}$ & 128	& $1\mathrm{e}^{-3}$	& 128 \\ 
& \small{Splice sites all} & $1\mathrm{e}^{-3}$ & 512	& $1\mathrm{e}^{-3}$	& 512 \\ 
& \small{Splice sites donors} & $1\mathrm{e}^{-3}$ & 128	& $1\mathrm{e}^{-3}$	& 128 \\ 
\bottomrule
\end{tabular}
\end{sc}
\end{small}
\end{center}
\end{table*}
%%%%%%%%%%%%%%%%%%%%%%%%%%%%%%%%%%%%%%%%%%%%%%%%%%%%%%%%%%%%%%%%%%%%%%%%%%%%%%%
%%%%%%%%%%%%%%%%%%%%%%%%%%%%%%%%%%%%%%%%%%%%%%%%%%%%%%%%%%%%%%%%%%%%%%%%%%%%%%%

\subsection{Predicting the Effect of Variants on Gene Expression}\label{appsubsec:vep}
Labels for this task represent whether a SNP has a causal effect on gene expression.
A positive label is assigned if the causal probability, as determined by the SuSiE \citep{wang2020simple} tool, is $> .9$ (see \citet{avsec2021effective}, where this task was originally proposed, for more details).
Chromosomes 9 and 10 are used as the held out test set (see \citet{trop2023LRB} for more details).

We follow the methodology presented in \citet{trop2023LRB} and extract embeddings for each model by taking an average of a 1536 bp window centered at the SNP location for both reference and alternative sequences and concatenating along the channel dimension.
Based on the tokenization scheme, for each model this window corresponds to a different number of tokens.
Namely, for HyenaDNA and Caduceus models, since base-pair-tokenization was used, the window consists of 1536 tokens as well.
Since Nucleotide Transformer was trained using 6-mer tokenization, the window corresponds to 256 bps.
Finally, for Enformer, the final embedding has a `receptive field' of 128 bps, hence a window of 12 `tokens' / positions is used.
To each embedding we also concatenate the tissue from which the sequence was assayed.

We also use a different input sequence length for each model.
For Caduceus and Hyena models, we use inputs of length 131k bps.
For Nucleotide Transformer, we use inputs of length 12k bps, which correspond to the input length on which this model was originally trained.
For Enformer, we use inputs of 196k bps, which correspond to the input length on which this model was originally trained.

We then train an SVM classifier with an RBF kernel on these embeddings for each strata of the data, which is separated by distance to nearest TSS.
For each bucket of distance to TSS, we randomly select 5,000 training points, fit an SVM with RBF kernel classifier, and record test set AUROC.
We repeat this process five times and report mean and +/- of one standard deviation across seeds.

Hyperparameter optimization was performed for each model within each distance category, focusing on the regularization strength.
We select this hyperparameter based on highest mean AUROC reported from 5 random seeds. 
The regularization strength used for each model reported in \Figref{fig:variant_effect_prediction} are listed in \Tabref{tab:variant_prediction_effect_hyperparam}.

Pre-trained weights for HyenaDNA were downloaded from \url{https://huggingface.co/LongSafari/hyenadna-medium-160k-seqlen-hf}.
Pre-trained weights for Nucleotide Transformer were downloaded from \url{https://huggingface.co/InstaDeepAI/nucleotide-transformer-v2-500m-multi-species}.
Pre-trained weights for the Enformer model were downloaded from \url{https://huggingface.co/EleutherAI/enformer-official-rough}.

\begin{table*}[th]
\caption{Hyperparameter Selection for SVM classifier in variant effect prediction task. Inverse of the $L_2$ regularization weight selected from $\{1,5,10\}$ by evaluating average test set AUROC.}
\label{tab:variant_prediction_effect_hyperparam}
\begin{center}
\begin{small}
\begin{sc}
\begin{tabular}{lccccccc}
\toprule
& \multicolumn{3}{c}{Distance to Nearest TSS (bp)} \\ 
& $0 - 30$k & $30 - 100$k & $100$k+ \\
\midrule
\small{Enformer} & 1 & 1 & 5 \\
\small{NTv2} & 1 & 1 & 10 \\ 
\small{HyenaDNA} & 1 & 1 & 5 \\ 
\small{Caduceus w/o Equiv} & 1 & 1 & 10 \\ 
\small{Caduceus-Ph} & 1 & 5 & 10 \\ 
\small{Caduceus-PS} & 1 & 1 & 5 \\ 
\bottomrule
\end{tabular}
\end{sc}
\end{small}
\end{center}
\end{table*}
\label{tab:sup_vep}

\section{Assets}\label{appsec:assets}
\subsection{Datasets}
For pre-training we use the HG38 human reference genome \citep{genome2009genome}.
The Genomics Benchmark comes from \citet{grevsova2023genomic}.
The Nucleotide Transformers benchmark is introduced in \citet{dalla2023nucleotide}.
The variant effect prediction task data was originally proposed in \citet{avsec2021effective} and we use the modified version from \citet{trop2023LRB}.

\subsection{Software and Libraries}
In \Tabref{tab:assets}, we enumerate the relevant open-source software, and corresponding licenses, used in this work.

\begin{table}[h]
    \caption{Open source libraries used in this work, with   corresponding licenses.}
    \label{tab:assets}
    \vskip 0.15in
    \begin{center}
    \begin{small}
    \begin{sc}
    \begin{tabular}{ll}
        \toprule
        Library & License \\
        \midrule
        GenomicsBenchmark~\citep{grevsova2023genomic} & Apache 2.0 \\
        Enformer PyTorch& MIT \\
        Mamba~\citep{gu2023mamba} & Apache 2.0 \\
        HuggingFace~\citep{wolf2019huggingface} & Apache 2.0 \\
        Hydra~\citep{Yadan2019Hydra} & MIT \\
        HyenaDNA~\citep{nguyen2023hyenadna} & Apache 2.0 \\
        NumPy~\citep{harris2020array} & \href{https://numpy.org/doc/stable/license.html}{NumPy license} \\
        Matplotlib~\citep{Hunter:2007} & \href{https://matplotlib.org/stable/users/project/license.html}{Matplotib license} \\
        ML Collections & Apache 2.0 \\
        OmegaConf & BSD 3-Clause \\
        Pandas \citep{reback2020pandas} & BSD 3-Clause ``New" or ``Revised" \\        PyTorch~\citep{Paszke_PyTorch_An_Imperative_2019} & BSD-3 Clause \\
        PyTorch Lightning~\citep{Falcon_PyTorch_Lightning_2019} & Apache 2.0 \\
        Scikit-Learn~\citep{scikit-learn} & BSD 3-Clause\\
        Seaborn~\citep{Waskom2021} & BSD 3-Clause ``New" or ``Revised" \\        
        Triton~\citep{tillet2019triton} & MIT \\
        \bottomrule
        \end{tabular}
    \end{sc}
    \end{small}
    \end{center}
    % \vskip -0.1in
\end{table}

\section{Computational resources}
Model training and inference were run on GPUs with number of devices and machine type varying by model size during pre-training and downstream tasks.
We use 3090, A5000, A6000, V100, and A100 GPUs.
\end{document}